\documentclass[aps,prl,showpacs,twocolumn,superscriptaddress, floatfix]{revtex4-1}%

\usepackage[utf8]{inputenc}
\usepackage{epsfig}
\usepackage{url}
\usepackage{multirow}
\usepackage{capt-of} % for captions of subtable / subfigure

\usepackage{graphicx}
\usepackage{amsfonts}
\usepackage{amssymb}
\usepackage{amsmath}
\usepackage{verbatim}
\usepackage{float} % used to place figure with specifier [H]]
\usepackage{bm} % bold math
\usepackage{qcircuit}
\usepackage{listings}
\usepackage{mathtools}

\usepackage{color}
\definecolor{codegreen}{rgb}{0,0.6,0}
\definecolor{codegray}{rgb}{0.5,0.5,0.5}
\definecolor{codepurple}{rgb}{0.58,0,0.82}
\definecolor{backcolour}{rgb}{0.95,0.95,0.92}
 
\lstdefinestyle{mystyle}{
    backgroundcolor=\color{backcolour},   
    commentstyle=\color{codegreen},
    keywordstyle=\color{magenta},
    stringstyle=\color{codepurple},
    basicstyle=\footnotesize,
    breakatwhitespace=false,         
    breaklines=true,                 
    captionpos=b,                    
    keepspaces=true,                 
    showspaces=false,                
    showstringspaces=false,
    showtabs=false,                  
    tabsize=2
}
 
\usepackage{hyperref}
\hypersetup{  colorlinks=true, linkcolor=blue, citecolor=red, urlcolor=blue  }

\usepackage{physics}

\renewcommand{\bf}[1]{\mathbf{#1}}
\renewcommand{\cal}[1]{\mathcal{#1}}
\newcommand{\bb}[1]{\mathbb{#1}}

\newcommand{\ba}{\begin{eqnarray}}
\newcommand{\ea}{\end{eqnarray}}

\newcommand{\beginsupplement}{%
        \setcounter{table}{0}
        \renewcommand{\thetable}{S\arabic{table}}%
        \setcounter{figure}{0}
        \renewcommand{\thefigure}{S\arabic{figure}}%
        \setcounter{equation}{0}
        \renewcommand{\theequation}{S\arabic{equation}}%
     }   % set counter in the Supplimental Material to zero

\usepackage{amsthm}
\newtheorem{theorem}{Theorem}
\newtheorem{lemma}{Lemma}

\theoremstyle{definition}
\newtheorem{definition}{Definition}

\usepackage{txfonts}
\begin{document}

\title{Experimental Cryptographic Verification for Near-Term Quantum Cloud Computing}
%\title{Experimental Cryptographic Verification with Instantaneous Quantum Computation}
%\title{Experimental Verification of Quantum Advantages with Instantaneous Quantum Computation}
%\title{Experimental Verification of Computational Advantages with Instantaneous Quantum Computation}
%\title{Experimental Demonstration of a Verification Protocol for Instantaneous Quantum Computation}
%\title{Experimental Observation of Quantum Advantage for Instantaneous Quantum Computation}

\author{Xi Chen}
\thanks{These two authors contributed equally}
\affiliation{ Hefei National Laboratory for Physical Sciences at the Microscale and Department of Modern Physics, University of Science and Technology of China (USTC), Hefei 230026, China}
\affiliation{CAS Key Laboratory of Microscale Magnetic Resonance, USTC, Hefei 230026, China}
\affiliation{Synergetic Innovation Center of Quantum Information and Quantum Physics, USTC, Hefei 230026, China}

\author{Bin Cheng}
\thanks{These two authors contributed equally}
\affiliation{Institute for Quantum Science and Engineering, and Department of Physics, Southern University of Science and Technology, Shenzhen 518055, China}

\author{Zhaokai Li}
\affiliation{ Hefei National Laboratory for Physical Sciences at the Microscale and Department of Modern Physics, University of Science and Technology of China (USTC), Hefei 230026, China}
\affiliation{CAS Key Laboratory of Microscale Magnetic Resonance, USTC, Hefei 230026, China}
\affiliation{Synergetic Innovation Center of Quantum Information and Quantum Physics, USTC, Hefei 230026, China}

\author{Xinfang Nie}
\affiliation{ Hefei National Laboratory for Physical Sciences at the Microscale and Department of Modern Physics, University of Science and Technology of China (USTC), Hefei 230026, China}
\affiliation{CAS Key Laboratory of Microscale Magnetic Resonance, USTC, Hefei 230026, China}
\affiliation{Synergetic Innovation Center of Quantum Information and Quantum Physics, USTC, Hefei 230026, China}

\author{Nengkun Yu}
\email{nengkunyu@gmail.com}
\affiliation{Centre for Quantum Software and Information, School of Software, Faculty of Engineering and Information Technology, University of Technology Sydney, NSW, Australia}

\author{Man-Hong Yung}
\email{yung@sustech.edu.cn}
\affiliation{Institute for Quantum Science and Engineering, and Department of Physics, Southern University of Science and Technology, Shenzhen 518055, China}
\affiliation{Shenzhen Key Laboratory of Quantum Science and Engineering, Southern University of Science and Technology, Shenzhen 518055, China}
\affiliation{Central Research Institute, Huawei Technologies, Shenzhen, 518129, P. R. China}

\author{Xinhua Peng}
\email{xhpeng@ustc.edu.cn}
\affiliation{ Hefei National Laboratory for Physical Sciences at the Microscale and Department of Modern Physics, University of Science and Technology of China (USTC), Hefei 230026, China}
\affiliation{CAS Key Laboratory of Microscale Magnetic Resonance, USTC, Hefei 230026, China}
\affiliation{Synergetic Innovation Center of Quantum Information and Quantum Physics, USTC, Hefei 230026, China}

% Include the date command, but leave its argument blank.

%\date{}

%\begin{CJK*}{GBK}{kai}

% Double-space the manuscript.

%\baselineskip24pt

% Make the title.

\begin{abstract}
% An important task for quantum cloud computing is to make sure that there is a real quantum computer running, instead of classical simulation. Here we explore the applicability of a cryptographic verification scheme for verifying quantum cloud computing. We provided a theoretical extension and implemented the scheme on a 5-qubit NMR quantum processor in the laboratory and a 5-qubit and 16-qubit processors of the IBM quantum cloud. We found that the experimental results of the NMR processor can be verified by the scheme with about $1.4\%$ error, after noise compensation by standard techniques. However, the fidelity of the IBM quantum cloud is currently too low to pass the test (about $42\%$ error). This verification scheme shall become practical when servers claim to offer quantum-computing resources that can achieve quantum supremacy.
% Required <= 600 characters, now 710
Recently, there are more and more organizations offering quantum-cloud services, where any client can access a quantum computer remotely through the internet. In the near future, these cloud servers may claim to offer quantum computing power out of reach of classical devices. An important task is to make sure that there is a real quantum computer running, instead of a simulation by a classical device. Here we explore the applicability of a cryptographic verification scheme that avoids the need of implementing a full quantum algorithm or requiring the clients to communicate with quantum resources. In this scheme, the client encodes a secret string in a scrambled IQP (instantaneous quantum polynomial) circuit sent to the quantum cloud in the form of classical message, and verify the computation by checking the probability bias of a class of output strings generated by the server. We provided a theoretical extension and implemented the scheme on a 5-qubit NMR quantum processor in the laboratory and a 5-qubit and 16-qubit processors of the IBM quantum cloud. We found that the experimental results of the NMR processor can be verified by the scheme with about $2.5\%$ error, after noise compensation by standard techniques. However, the fidelity of the IBM quantum cloud is currently too low to pass the test (about $42\%$ error). This verification scheme shall become practical when servers claim to offer quantum-computing resources that can achieve quantum supremacy. 
% \red{>227 words}
\end{abstract}

\maketitle

%%%%%%%%%%%%%%%%%%%%%%%%%%%%%%%%%%%%%%%%%%%%%%%%

%%%%%%Introduction

\textit{Introduction.---} 
Quantum computation promises a regime with unprecedented computational power over classical devices, offering numerous interesting applications, such as factorization~\cite{Shor-factorization}, quantum simulation~\cite{Feynman1982-simulation, Lloyd1996-simulation}, and quantum machine learning~\cite{Harrow2009-HHL,Lloyd13-supervised-and-unsupervised}. However, before quantum computers become prevalent to the public, one might expect that only organizations with sufficient resources could operate a full-scale quantum computer, analogous to today's supercomputers. Furthermore, individuals who have demands for quantum computation could access the service through the internet, i.e., cloud quantum computing. In fact, several small-scale quantum cloud services have already been launched~\cite{ibmqx, rigetti, bristol}, %\red{check the active ones} 
which can be operated by remote clients through the internet. As a result, many simulations performed from quantum cloud servers have been reported (see Ref.~\cite{1804.03719-tutorial-beginner} for a summary).

In the near future, it is not impossible that these clouds may claim to offer 100 or more working qubits and many layers of quantum gates, where quantum supremacy~\cite{Preskill2012,Lund2017,Terhal2018,YungNSR} could be achieved. However, one may naturally ask, is there a real quantum computer behind the cloud? Or, would it just be a classical computer simulating quantum computation? For ordinary clients who only have control and access of classical computer, a natural task is to verify whether these cloud servers are truly quantum. 

Alternatively, the question can be formalized as follows: {\it is it possible for a purely-classical client to verify the output of a quantum prover?} This question has been extensively explored for more than ten years. In 2004, Gottesman initialized this question, which Aaronson wrote down in his blog \cite{Gottesman04}. 
A straight-forward idea is to run a quantum algorithm solving certain NP problems, for example, Shor's algorithm for integer factorization~\cite{Shor-factorization}. Such problems might be hard for classical computation, but are easy for \emph{classical verification} once the result is known.
However, the challenge is that a full quantum algorithm typically requires thousands of qubits and quantum error correction to be implemented, which is out of question in the NISQ~\cite{Preskill2018} (Noisy Intermediate-Scale Quantum)  era. 

Note that the verification problem have different variants. For example, one may assume that the supposedly ``classical" client may actually have a limited ability to perform quantum operations on a small number of qubits. This line of research has already attracted much attention ~\cite{Broadbent2008,Broadbent2010-blind-original,Aharonov2008,Aharonov2017, Fitzsimons2017-verifiable-blind,Fitzsimons2018, Mills2018-it-secure}. Without any quantum power, the client might still be able to verify delegated quantum computation which is spatially separated and entanglement can be shared~\cite{Reichardt2013-blind-vazirani, Huang2017-blind-experiment}. Currently, this approach does not seem to fit the setting of the available quantum cloud services, but it does reveal the outstanding challenge for establishing a rigorous verification scheme based on a classical client interacting with a single server using only classical communication~\cite{Aaronson2017-blind-implausibility}. 

Until recently, Mahadev has made important progresses~\cite{Mahadev2017,Mahadev2018}, assuming that the learning-with-errors problem~\cite{regev2009lwe} is computationally hard even for quantum computer. The protocol allows a classical computer to {\it interactively} verify the results of an efficient quantum computation, achieving a fully-homomorphic encryption scheme for quantum circuits with classical keys. Despite these great efforts, we are still facing the problems of ``non-interactively" verifying {\it near-term} quantum clouds, which would be too noisy for implementing full quantum algorithms but may be capable of demonstrating quantum supremacy.

Here we report an experimental demonstration of a simple but powerful cryptographic verification protocol, originally proposed by Bremner and Shepherd~\cite{IQP08} in 2008. We extended the theoretical construction in terms of $n$-point correlation. The implementation was first performed with a 5-qubit NMR quantum processor in the laboratory. Additionally, we also benchmarked the performance of the verification scheme by actually implementing the protocol with the IBM quantum cloud processors~\cite{ibmqx}.

The verification protocol implemented is based on a simplified circuit model of quantum computation, called  IQP (instantaneous quantum polynomial) model~\cite{IQP08}; the qubits are always initialized in the `0' state. The IQP circuits contain three parts. In the first and the last part, single-qubit Hadamard gates are applied to every qubit. The middle part of an IQP circuit does not contain an explicit  temporal structure, in the sense that diagonal (and hence commuting) gates acting on single or multiple qubits are applied. On one hand, the IQP model  represents a relatively resource-friendly computational model to be tested with near-term quantum devices. On the other hand, the IQP model has been proven to be hard for classical simulation~\cite{IQP10,IQP15}, under certain computational assumptions, similar to Boson sampling~\cite{Aaronson2010}. 

\begin{figure}[t]
    \centering
    \includegraphics[width = 0.9\columnwidth]{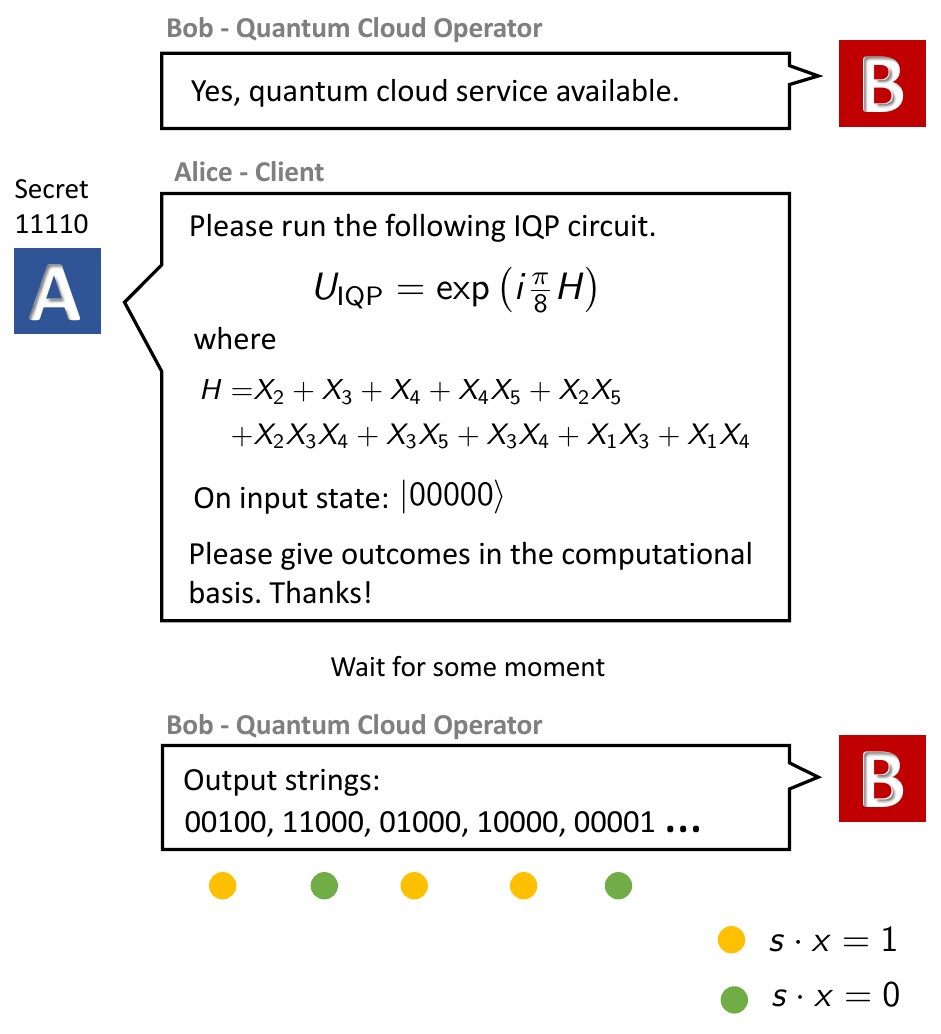}
    \caption{Schematic representation of the protocol. Alice generates a description matrix, according to the construction of quadratic residue code. This description matrix has an associating secret vector $s$ and determines an $X$-program circuit. Bob runs this circuit, measures and sends back the measurement data to Alice. From Bob's data, Alice computes the probability bias ${\mathcal{P}_{s \bot }}$, with respect to the secret vector $s$, and sees whether it is close to 0.854, to decide whether Bob has a true quantum device or not.}
    \label{fig:protocol}
\end{figure}

\textit{Verification protocol.---}
In the cryptographic verification protocol~\cite{IQP08}, there are two parties labeled as Alice (the client) and Bob (the server). Alice is assumed to be {\it completely classical}; she can only communicate with others through classical communication (e.g., internet).  Suppose Bob claims to own a quantum computer and Alice is going to test it. In reality, of course, there is no need for Alice to inform Bob about her intention; she may just pretend to run a normal quantum program. The protocol can be succinctly summarized as follows (depicted by Fig.~\ref{fig:protocol}).
\begin{enumerate}
  \item[{\textbf{Step 1:}}] Alice first generates a matrix (called X-program~\cite{IQP08}) associated with a secret string $s \in \{0,1\}^n$, which is only kept by Alice.
  \item[ {\textbf{Step 2:}}] Alice then translates the X-program into an IQP circuit of $n$ qubits, and sends the information about the IQP circuit $U_{\rm IQP}$ to Bob.
  \item[{\textbf{Step 3:}}] Bob returns the outputs to Alice in terms of the bit strings $x \in \{0,1\}^n$, which should follow the distribution of the IQP circuit, i.e., $\Pr \left( x \right) = {| {\left\langle x \right|{U_{{\text{IQP}}}}\left| {{0^n}} \right\rangle } |^2}$, if Bob is honest.
   \item[{\textbf{Step 4:}}] Ideally, Alice should be able to determine if the probability distributions $\Pr(x)$ for a subset of strings orthogonal to the secret string, where $x \cdot s \equiv {x_1}{s_1} + {x_2}{s_2} + \cdots + {x_n}{s_n} = 0 {\mod 2}$, add up to an expected value 0.854. Otherwise, Bob fails to pass the test.
\end{enumerate}

More specifically, the key quantity of interest is the following probability bias defined by, 
\begin{align}\label{eq:bias}
{\mathcal{P}_{s \bot }} \equiv \sum\limits_{x \in {{\left\{ {0,1} \right\}}^n}} {|\langle x|{U_{{\text{IQP}}}}\left| {{0^n}} \right\rangle {|^2}} \;{\delta _{x \cdot s = 0}} \ ,
\end{align}
where ${\delta _{x \cdot s = 0}} = 1$ if it is true that ${x \cdot s = 0}$, and ${\delta _{x \cdot s = 0}} = 0$ otherwise. For a perfect quantum computation, the value of the probability bias should be ${\mathcal{P}_{s \bot} } = 0.854$. The best known classical algorithm~\cite{IQP08} would instead produce a value of 0.75, which is relevant when $n$ is sufficiently large. This quantum-classical gap in the probability bias makes it possible to apply such a resource-friendly cryptographic verification scheme for testing quantum cloud computing in the regime where quantum supremacy would be achieved.

% The IQP circuit run in our experiments is shown in Fig.~\ref{fig:X_program} and the secret vector is $s=(1,1,1,1,0)$. The method to generate the IQP circuit (or X-program matrix) and the associating secret vector is described in the Discussion section. 

\begin{figure*}[t]
    \centering
    \includegraphics[width = 1.8\columnwidth]{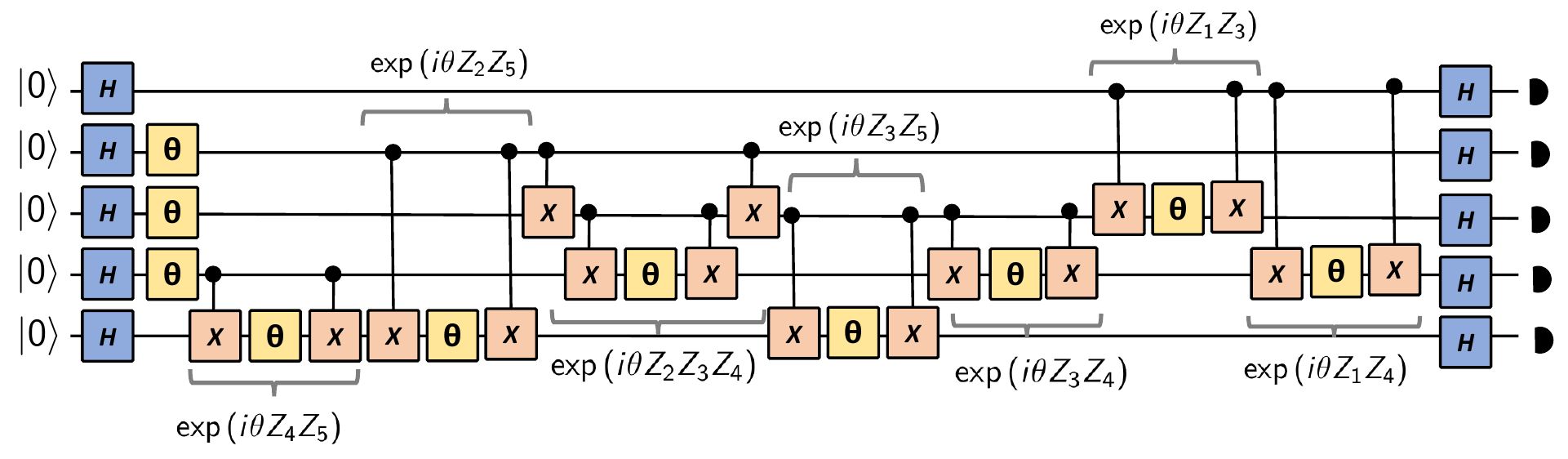}
    \caption{IQP circuit for the X-program matrix~\eqref{eq:matrix}. There are two layers of Hadamard gates at the beginning and the end of the circuit. The gates in between are all diagonal in the $Z$ basis. $\theta$ stands for $\exp(i\theta Z)$.
    }
    \label{fig:X_program}
\end{figure*}

\textit{Overview.---}
To illustrate our experimental demonstrations, we shall first provide a concise and self-contained theoretical description of the cryptographic protocol. Particularly, our computer code implemented in the experiment is open source and available online~\cite{gitlab} and in Supplemental Materials; interested readers can readily reproduce our results with it, and can also apply it to generate X-programs of different variations for testing other quantum-cloud services. 

On the other hand, we also provide a theoretical extension of the original work~\cite{IQP08}, transforming it into a form more familiar to the physics community. Specifically, we connect the probability bias in Eq.~(\ref{eq:bias}) with the Fourier coefficient of the probability $\Pr(x)$ of the output strings. As a result, we can express the probability bias through the $n$-point correlation function (see Supplemental Materials):
\begin{equation}\label{reprePs}
{\mathcal{P}_{s \bot }} = \frac{1}{2}\left( {1 + \left\langle {{Z^{{s_1}}}{Z^{{s_2}}} \cdots {Z^{{s_n}}}} \right\rangle } \right) \ .
\end{equation}
Since the string $s = {s_1}{s_2} \cdots {s_n}$ is not known to Bob, the verification protocol can be regarded as a game where Alice tests the outcomes in terms of a particular correlation function unknown to Bob.

In addition, as we will see later, the X-program consists of two part, the main part and the redundant parts, and the representation of Eq.~(\ref{reprePs}) provides a straight-forward way to understand why the redundant part of the X-program does not affect the probability bias---they commute with the $n$-point correlation function. 

Our theoretical extension in Eq.~(\ref{reprePs}) allows us to take into account the effect of noises. More precisely, if one models~\cite{Bremner2016,Yung2017} the decoherence by a dephasing channel (with an error rate $\epsilon$) applied for each qubit at each time step, then the probability bias becomes ${\mathcal{P}_{s \bot }} \to \frac{1}{2}\left( {1 + {{\left( {1 - 2 \epsilon } \right)}^{\left| s \right|}}\left\langle {{Z^{{s_1}}}{Z^{{s_2}}} \cdots {Z^{{s_n}}}} \right\rangle } \right)$, where ${\left| s \right|}$ is the Hamming weight of $s$, that is the number of 1's in $s$.

Experimentally, our data were taken separately from two different sources, namely a five-qubit NMR processor in the laboratory, and the IBM cloud services, aiming to benchmark the performances of the IQP circuit implementation under the laboratory conditions and that from the quantum-cloud service. 

Our results show that the laboratory NMR quantum processor can be employed to verify the IQP circuit after noise compensation by standard techniques, but the IBM quantum cloud was too noisy. The probability bias obtained from the IBM's processors are close to 0.5, which is the result of uniform distribution. The main reason is that IBM's system has many constraints on the connectivity between the physical qubits; we had to include many extra SWAP gates to complete the circuit, causing a severe decoherence problem. 

\textit{Theoretical construction.---}  
%%%%%%%%%%%%%%%%%%
Here Alice's secret vector of string is given by $s = (1,1,1,1,0)$. An X-program can be represented by a matrix with binary values, which is constructed from the quadratic residue code (QRC)~\cite{macwilliams1977book}. In the experiment, the matrix $\mathcal{X}$ associated with the X-program is given by the following (see Supplemental Material for the construction method),
%\begin{align}\label{eq:matrix}
%\mathcal{X}_{\rm pro} = 
%\begin{pmatrix}
%0&0&0&0&0&0&0&0&1&1 \\
%1&0&0&0&1&1&0&0&0&0 \\
%0&1&0&0&0&1&1&1&1&0 \\
%0&0&1&1&0&1&0&1&0&1 \\
%0&0&0&1&1&0&1&0&0&0
%\end{pmatrix}^T \ .
%\end{align}
\begin{align}\label{eq:matrix}
{\mathcal{X}} = {\left( {\begin{array}{*{20}{c}}
  {\underbrace {\begin{array}{*{20}{c}}
  0&0&0&0&0&0&0 \\ 
  1&0&0&0&1&1&0 \\ 
  0&1&0&0&0&1&1 \\ 
  0&0&1&1&0&1&0 \\ 
  0&0&0&1&1&0&1 
\end{array}}_{{\text{Main}}}}&{\underbrace {\begin{array}{*{20}{c}}
  0&1&1 \\ 
  0&0&0 \\ 
  1&1&0 \\ 
  1&0&1 \\ 
  0&0&0 
\end{array}}_{{\text{Redundant}}}} 
\end{array}} \right)} \ .
\end{align}

Here the X-program has a layer of security for protecting the knowledge of the secret string $s$ from Bob. Explicitly, there are two parts in the matrix, (i) the main part and (ii) a redundant part. Columns in the main part are not orthogonal to the secret vector $s$, i.e., $x \cdot s = 1$, while columns in the redundant part are, i.e., $x \cdot s = 0$. Both parts have to be changed if the secret string is changed. 

However, an important property of the X-program is that the probability bias ${\mathcal{P}_{s \bot} }$ depends only on the main part. So Alice can append as many redundant columns that are orthogonal to $s$ to this matrix as she wishes. Of course, later she would need to scramble the columns, in order to hide the secret $s$ from Bob.  

Next, the X-program has to be translated into an IQP circuit~\cite{IQP08}, which is a subclass of quantum circuits with commuting gates before and after the Hadamard gates.  Equivalently, the unitary transformation associated with the IQP circuit can be casted as follows: 
\begin{equation}\label{eq:U_IQP}
{U_{{\text{IQP}}}} = \exp \left( {i\theta H} \right) \ ,
\end{equation}
where $\theta = \pi/8$, and $H$ is the effective Hamiltonian constructed by the elements of the X-program. For example, a column $(1,0,1,1,0)$ represents a term $X_1 X_3 X_4$, where $X_i$ is a Pauli-$X$ acting on the $i$-th qubit. As a result, the full Hamiltonian corresponding to $\cal{X}$ reads,
\begin{align}
H &= X_2 + X_3 + X_4 + X_4X_5 + X_2X_5 \notag \\
&+ X_2X_3X_4 + X_3X_5 + X_3X_4 + X_1X_3 + X_1X_4 \ . \label{eq:Hamiltonian} 
\end{align}
Note that if we take $\theta = \pi/4$, then the evolution can be simulated classically by the Gottesman-Knill algorithm~\cite{Gottesman98}. Fig.~\ref{fig:X_program} shows the circuit diagram.

In the ideal case, the probability bias for the IQP circuit should be given by ${\mathcal{P}_{s \bot }} = 0.854$. If Bob outputs random bits, the value of the probability bias would be ${\mathcal{P}_{s \bot }} = 0.5$. However, although it is scrambled, the X-program is correlated with the secret string. The classical algorithm provided in Ref.~\cite{IQP08} can yield ${\mathcal{P}_{s \bot }} = 0.75$, which was conjectured to be optimal~\cite{IQP08}. As a result, it becomes possible to verify the quantum hardware behind the quantum clouds by simply collecting the statistics of the outputs to check if we can get ${\mathcal{P}_{s \bot }} = 0.854$.

For the purpose of benchmarking, we performed a total of three separate implementations of the same X-program on an NMR quantum processor and on IBM quantum processors, including the 5-qubit one and the 16-qubit one~\cite{ibmqx}.

\begin{figure}[t!]
    \centering
    \includegraphics[width = 0.9\columnwidth]{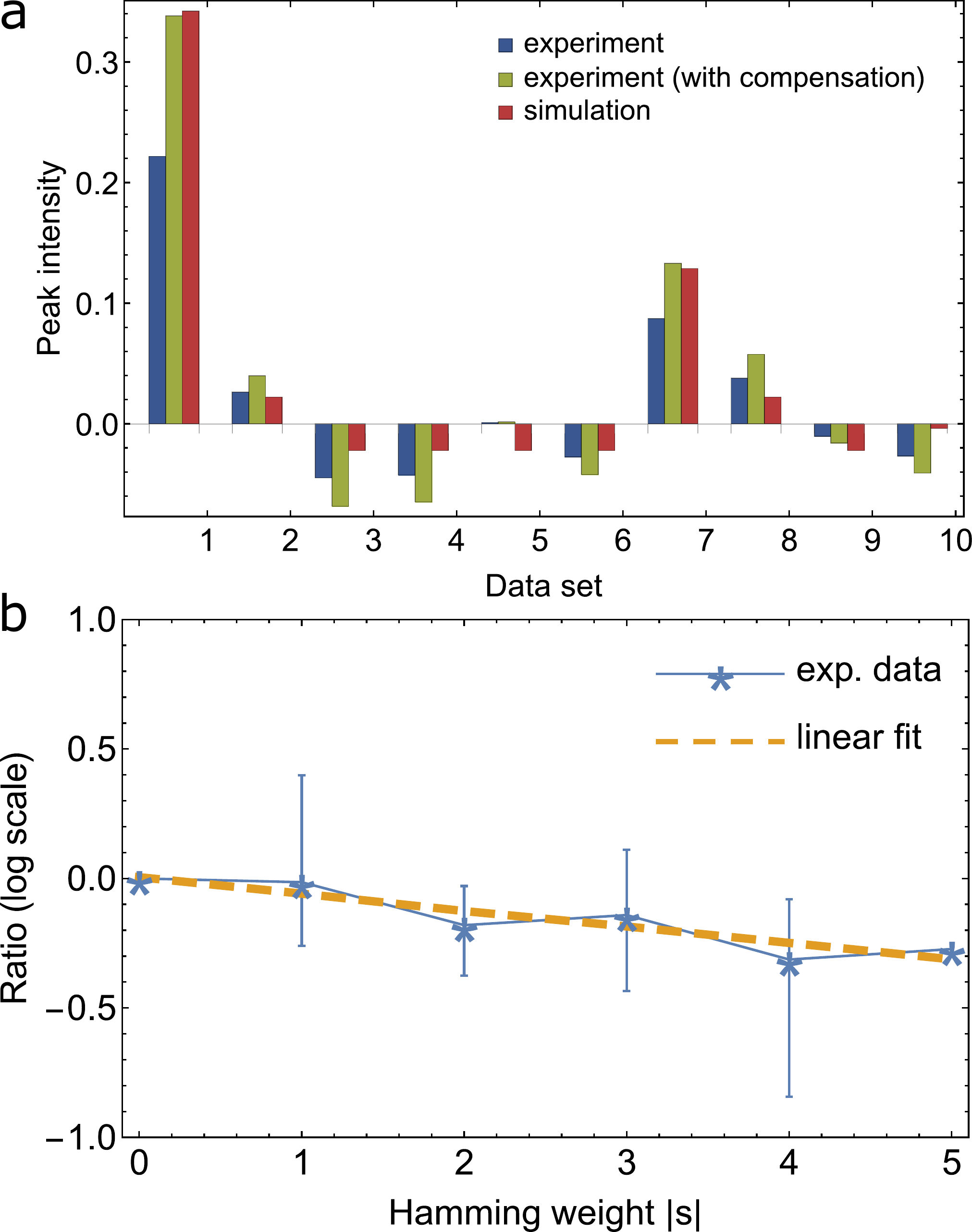}
    \caption{\textbf{(a)} Part of the peak intensities from the readout pulses. There are 80 groups of peak intensities in total and we only present the first 10 groups here. Each peak intensity is a linear combination of probabilities. \textbf{(b)} The ratio of the experimental value to the theoretical value of the $n$-point correlation function $\langle Z^{s_1} Z^{s_2} \cdots Z^{s_n} \rangle$ in log scale, versus Hamming weights.
     }
    \label{fig:exp_data}
\end{figure}

\begin{figure}[t!]
    \centering
    \includegraphics[width = 0.9\columnwidth]{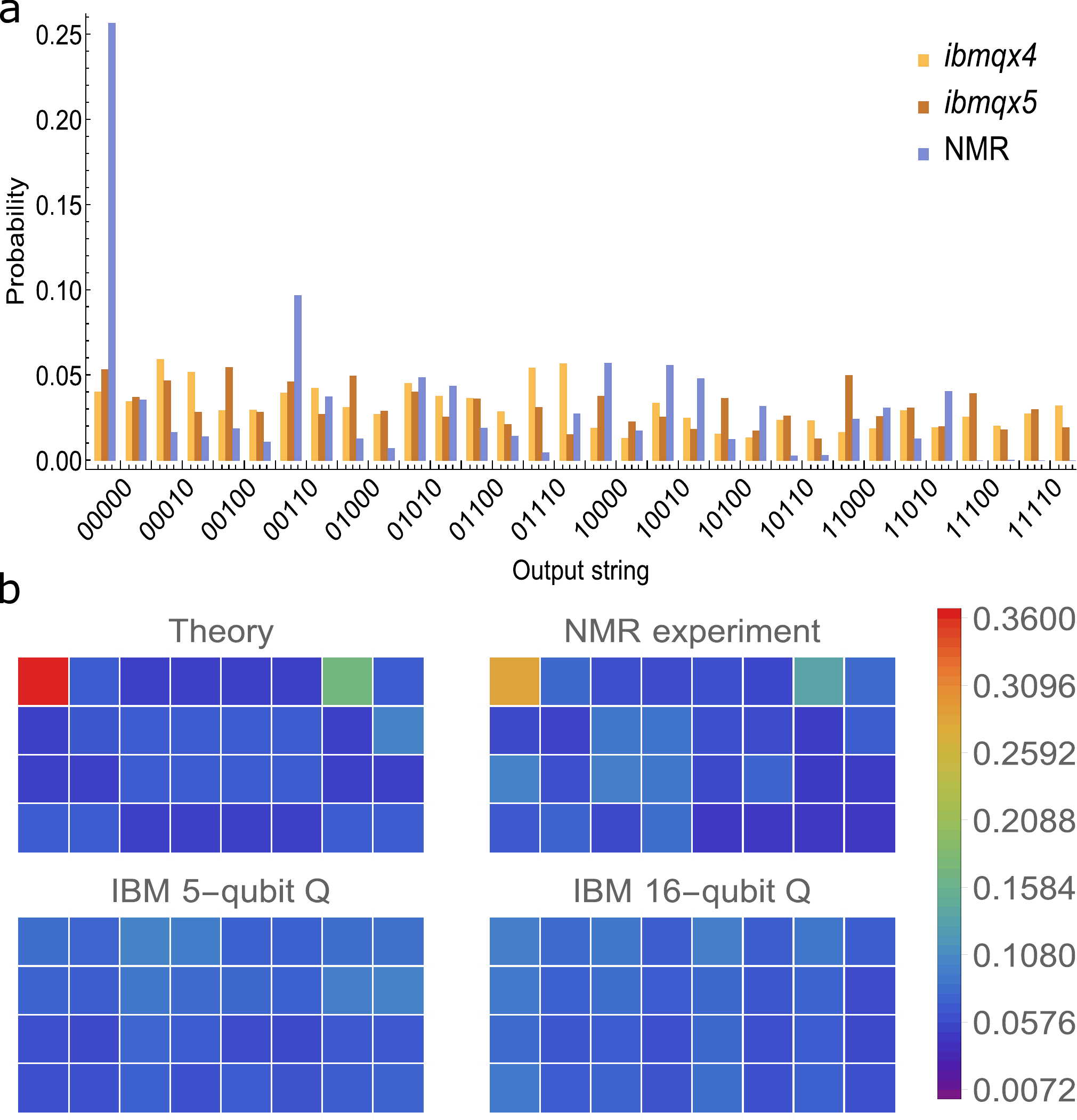}
    \caption{\textbf{(a)} Probability distributions from IBM quantum processors and the NMR processor. \emph{ibmqx4} is the 5-qubit processor and \emph{ibmqx5} is the 16-qubit one. \textbf{(b)} The probabilities are put into grids and the colors indicate their values according to the color scale on the right. 
     }
    \label{fig:NMR_vs_ibm}
\end{figure}

\textit{Verification with NMR in the laboratory.---}
The experiments with the NMR quantum processor were carried on a Bruker AV-400 spectrometer at 303K. The 5-qubit quantum processor consists of two ${}^1H$ nuclear spins and three ${}^{19}F$ nuclear spins in 1-bromo-2,4,5-trifluorobenzene dissolved in the liquid crystal N-(4-methoxybenzylidene)-4-butylaniline (MBBA)~\cite{SHANKAR201410-sample}. 
The molecular structure and equilibrium spectra of ${}^{19}F$ nuclear spins are provided in the Supplemental Materials.

Starting from the thermal state $\rho_{\rm eq}$, the NMR system is initially prepared in a PPS ${\rho _i} = \frac{{1 - \varepsilon }}{{32}}\mathbf{I}_{32} + \varepsilon \dyad{00000}$ by the line-selective method~\cite{Peng2001-line-selective}. Here $\mathbf{I}_{32}$ represents the $32 \times 32$ identity operator and $\varepsilon  \approx {10^{ - 5}}$ is the polarization. Note that the identity operator is invariant under the unitary transformation, neither does it affect the measurement step. The state $\rho_i$ evolves the same way as a true pure state $\dyad{00000}$ and generates the same signal up to a proportionality factor $\varepsilon$, so we can simply regard the ${\rho _i} $ as $\dyad{00000}$.

% Then we apply the unitary transformation $U_{\rm IQP}$ as well as five readout pulses to the initial state $\rho_i$.
% In the actual implementation, we packed the quantum circuit $U_{\rm IQP}$ with the five readout pulses together into five shape pulses optimized by the gradient ascent pulse engineering (GRAPE) method~\cite{KHANEJA2005296-grape}, with the length of each shape pulse being 37.5 ms and the number of segments being 1500. All the pulses have theoretical fidelity over $99.4\%$ and are designed to be robust against the inhomogeneity of the pulse amplitude.

To implement the IQP circuit, i.e. the $U_{\rm IQP}$ described by Eq.~(\ref{eq:U_IQP}),
we packed it into one shaped pulse optimized by the gradient ascent pulse engineering (GRAPE) method~\cite{KHANEJA2005296-grape}, with the length being 37.5 ms and the number of segments being 1500. The shaped pulse has the theoretical fidelity of $99.4\%$ and is designed to be robust against the inhomogeneity of the pulse amplitude.

% In the NMR system, we first prepare a pseudo-pure state (PPS), which is effectively a $\ket{00000}$ state. After that, we implement the unitary transformation $U_{\rm IQP}$ described by Eq.~\eqref{eq:U_IQP}, followed by the readout pulses. 
To obtain the probability bias in the experiment, we need five readout pulses (i.e. a $Y$ pulse on each qubit $\exp(-i\pi Y_j /4)$) to reconstruct the diagonal elements of the density matrix of the final state~\cite{tomography}, which are the probabilities in the computational basis. For details, we refer to the Supplemental Materials.

From each readout pulse, we can obtain 16 peak intensities, and each peak intensity is a linear combination of the 32 probabilities. So we have 80 linear equations of the form: $\sum_{x} c_x(l) \Pr(x) = \alpha_l$ ($1\leq l\leq 80$), where $\alpha_l$'s are the peak intensities read out by our device. We present 10 of these peak intensities in Fig.~\ref{fig:exp_data}~(a), and figure with all peak intensities can be found in Supplemental Materials. 
% In Fig.~\ref{fig:NMR_vs_ibm} (a), there are three different kinds of data in each group. 
The blue lines in Fig.~\ref{fig:exp_data}~(a) are from the experiment on our NMR processor, %the green one is from theoretical simulation with noise in our NMR processor taken into account, 
and the red ones are from theoretical simulation without considering the noise effect. After solving those 80 linear equations  from NMR processor together with the normalization condition $\sum_x \Pr(x) = 1$ through the least square method, we obtain the corresponding probability distribution, as depicted in Fig.~\ref{fig:NMR_vs_ibm}~(a) (the blue histogram). The probability bias from this raw distribution is 0.755.%0.711. 

The probability bias is connected to the $n$-point correlation function $\langle Z^{s_1} Z^{s_2} \cdots Z^{s_n} \rangle$ through Eq.~\eqref{reprePs}. If there is single-qubit dephasing noise on every qubits at each step, the $n$-point correlation function will decay by a factor $(1 - 2\epsilon)^{|s|}$~\cite{Bremner2016, Yung2017}. Thus the ratio of the experimental $n$-point correlation (which is from the raw distribution) to the theoretical value is $(1 - 2\epsilon)^{|s|}$. Fig.~(\ref{fig:exp_data})~(b) shows this ratio in log scale, versus Hamming weights of all possible $s$. The slope of the linear fit is $\log(1 - 2\epsilon)$, from which we obtain an effective noise rate $\epsilon = 6.79\%$.

We note that the whole duration of the dynamic evolution is 37.5 ms whereas the decoherence time is about 50 ms. Hence, the decay caused by the decoherence is not negligible. To compensate the effects of decoherence, we experimentally estimate the attenuation factor, that is, the ratio of peak intensities with decoherence to peak intensities without decoherence. Concretely, we design a shaped pulse of the $32\times 32$ identity operator with length 37.5 ms and use the decay in the peak intensities of this identity evolution to estimate the attenuation factor. 
To compensate the effects of decoherence, peak intensities from actual experiment are divided by the attenuation factor (see Supplemental Material for details) and the resulted peak intensities are shown as the green lines in Fig.~\ref{fig:exp_data}.
Then with a similar method by solving linear equations, we derive the probability bias compensated by noise: 0.866 $\pm$ 0.016 (comparable with the theoretical value: 0.854). Details of the error bar 0.016 can be found in Supplemental Material.

\textit{Verification with IBM cloud.---} 
As for experiments on IBM devices, we run the same circuits as in Fig.~\ref{fig:X_program}. However, due to the connectivity constraints~\cite{ibmqx}, we have to include several additional SWAPs to complete the circuit, which makes the whole circuit depth be about $40\sim 50$; for example, CNOT gates can only be applied to a certain pairs of qubits. In our demonstration, we applied the verification algorithm for \emph{ibmqx4}, a 5-qubit superconducting processor, and \emph{ibmqx5}, a 16-qubit one, which are accessed via a software called QISKit.  
%To implement the IQP circuits shown in  Fig.~\ref{fig:X_program}, we need to implement many SWAP gates and the whole circuit depth is about $40\sim 50$.
We collect specification of IBM devices in Supplemental Material, including the connectivity and coherence time. These data can also be found in Ref.~\cite{ibmqx}.

In Fig.~\ref{fig:NMR_vs_ibm}~(a), the histogram in orange is the probability distribution from the experiment on \emph{ibmqx4}, while the brown one is from \emph{ibmqx5}. The probability biases from these two distributions are respectively 0.488 and 0.492, which are far from the expected value of 0.854. In fact, from a completely-mixed state, we can obtain a probability bias of 0.5. So the values of bias from these two quantum cloud services by IBM indicate that their final states are highly corrupted by decoherence. Furthermore, to see whether the IBM cloud would have a better performance if we reduce the depth, we implement a quantum circuit only corresponding to the main part of matrix~\eqref{eq:matrix} on \emph{ibmqx4}. However, the obtained bias is 0.512, which is still very close to that from a completely mixed state (see Supplemental Material for details). Therefore, we concluded that the IBM cloud was too noisy to pass our test.

Fig.~\ref{fig:NMR_vs_ibm}~(b) shows the comparison of the distributions. Each grid has 32 elements, corresponding to 32 probabilities. The color in each element indicates the concrete value, according to the color scale on the right. The distribution from the experiment run on NMR device is close to the theoretical prediction while the last two, which are distributions from IBM devices, are not.

\textit{Summary.---}
In conclusion, we have performed a proof-of-principle demonstration of a cryptographic verification scheme, using an NMR quantum processor and the IBM quantum cloud. The experimental results show that the fidelity of the quantum cloud service has to be significantly improved, in order to be testable with the verification method. In particular, the connectivity between the qubits imposes an extra overhead in the implementation of the scheme. For a large-scale implementation, it is the also important to determine numerically the size of IQP circuit that can no longer be simulable by classical computers, which is currently an open question.

% \section{Data availability}
% The authors declare that all data supporting the findings of this study are available within the article and its Supplementary Information files, or are available from the authors upon request.

\textit{Acknowledgement.---}
We acknowledge use of the IBM Q for this work. The views expressed are those of the authors and do not reflect the official policy or position of IBM or the IBM Q team. 
XP, XC, ZL and XN are supported by the National Key Research and Development Program of China (Grant No. 2018YFA0306600), the National Science Fund for Distinguished Young Scholars (Grant No. 11425523), the National Natural Science Foundation of China (Grants No. 11575173), Projects of International Cooperation and Exchanges NSFC (Grant No. 11661161018), and Anhui Initiative in Quantum Information Technologies (Grant No. AHY050000).
NY is supported in part by the Australian Research Council (Grant No. DE180100156). 
MHY is supported by the National Natural Science Foundation of China (11875160), the Guangdong Innovative and Entrepreneurial Research Team Program (2016ZT06D348), Natural Science Foundation of Guangdong Province (2017B030308003), and Science, Technology and Innovation Commission of Shenzhen Municipality (ZDSYS20170303165926217, JCYJ20170412152620376, JCYJ20170817105046702).

%\nocite{*}
% \bibliographystyle{unsrt} % full title in ref
\bibliography{ref}

% \textbf{Author contributions:}
% M.H.Y and X.P designed research; N.Y, X.C and B.C performed research; Z.L and X.N helped analyze data; and X.C, B.C, Z.L, N.Y, M.H.Y and X.P wrote the paper.

% \textbf{Competing interest: } The authors declare no competing interests.

% Natural Science Foundation of Guangdong Province (2017B030308003) and the Guangdong Innovative and Entrepreneurial Research Team Program (No. 2016ZT06D348), and the Science Technology and Innovation Commission of Shenzhen Municipality (ZDSYS20170303165926217, JCYJ20170412152620376). 
% XP, XC and XN acknowledge the support from the National Key Research and Development Program of China (2018YFA0306600), the National Key Basic Research Program of China (2014CB848700), National Natural Science Foundation of China (grants No. 11425523, 11375167, 11661161018, and 11227901), and Anhui Initiative in Quantum Information Technologies(grant No. AHY050000), and the Postdoctoral Science Foundation of China (No. 2018M632195).

\newpage

\clearpage

% ------------------------------------------------------------------------------------------------------------------------------------------------------------------------------------------------------------------------------------------------
% ---------------------------------------------------------------------------------------------------------------  APPENDIX ----------------------------------------------------------------------------------------------------------------
% ------------------------------------------------------------------------------------------------------------------------------------------------------------------------------------------------------------------------------------------------
\newpage
\onecolumngrid
%\begin{appendix}

\section{Supplemental Material}

\beginsupplement

\subsection{Quadratic-Residue-Code Construction}

We briefly review the quadratic-residue-code (QRC) construction. For detailed proofs, we refer readers to Ref.~\cite{IQP08} and Ref.~\cite{macwilliams1977book}. Below, all linear algebraic objects (vectors, matrices, linear spaces, etc.) are over $\mathbb{F}_2$. 

\subsubsection{Related Classical Coding Theories}

\begin{definition}
\emph{(code and codeword)} A \emph{code} $\cal{C}$ is a linear subspace of $\bb{F}_2^n$, and the elements of a code is called \emph{codeword}. Denote ``$c$ is a codeword of $\cal{C}$'' as $c\in \cal{C}$.
\end{definition}
Given a matrix $\mathcal{X}$, the linear combination of its rows generate a code $\mathcal{C}$, and such a matrix is called \emph{generating matrix} of $\cal{C}$. Generating matrices for a code are not unique. 
\begin{definition}
\emph{(quadratic residue)} An integer $j$ is called a quadratic residue modulo $q$ if there exists an integer $x$, \emph{s.t.} $x^2 \equiv j \text{ (mod } q)$.
\end{definition}
Suppose $q$ is a prime such that $8$ divides $q + 1$. Consider a binary vector of length $q$, the $j$-th components of which is 1 if and only if $j$ is a quadratic residue modulo $q$. The smallest example is $q = 7$, in which case such vector is $(1,1,0,1,0,0,0)$. Rotate it, and we get a class of vectors, such as $(0,1,1,0,1,0,0)$, $(0,0,1,1,0,1,0)$, etc.  This class of vectors generates a linear subspace $\cal{C}_{\rm QRC}$, called \emph{quadratic residue code} (QRC). For $q = 7$, the matrix below can be a generating matrix: 
\begin{align}
\cal{X}_M = 
\begin{pmatrix}
1 & 1 & 1 & 1 & 1 & 1 & 1 \\
1 & 1 & 0 & 1 & 0 & 0 & 0 \\
0 & 1 & 1 & 0 & 1 & 0 & 0 \\
0 & 0 & 1 & 1 & 0 & 1 & 0 \\
0 & 0 & 0 & 1 & 1 & 0 & 1 
\end{pmatrix} \equiv 
\begin{pmatrix}
r_0 \\
r_1 \\
r_2 \\
r_3 \\
r_4 \\
\end{pmatrix}\label{eq:main}
\end{align}
% \[
% P_s = 
% \begin{pmatrix}
% 1 & 1 & 0 & 0 & 0 \\
% 1 & 1 & 1 & 0 & 0 \\
% 1 & 0 & 1 & 1 & 0 \\
% 1 & 1 & 0 & 1 & 1 \\
% 1 & 0 & 1 & 0 & 1 \\
% 1 & 0 & 0 & 1 & 0 \\
% 1 & 0 & 0 & 0 & 1 
% \end{pmatrix} \ .
% \]
Here, the last $4$ rows form a basis, and the first row is just a linear combination of the basis, thereby leaving the code invariant. More explicitly, any vectors in the code space generated by $\mathcal{X}_M$ can be written as $\sum_{i = 1}^4 c_i r_i$, where $c_i \in \{ 0, 1 \}$.

\begin{definition}
\emph{(Hamming weight)} The \emph{Hamming weight} of a codeword is the number of $1$'s that it has, denoted as $|c|$.
\end{definition}
If the Hamming wieght is even, then we say the codeword is \emph{even}. 
\begin{definition}
\emph{(doubly even)} A code is \emph{doubly even} if all the codewords have Hamming weight a multiple of 4.
\end{definition}

\begin{definition}
\emph{(dual code and self-dual)} For a code $\cal{C} \subseteq \bb{F}_2^n$, the \emph{dual code} $\cal{C^{\perp}}$ is defined as $\cal{C^{\perp}} = \{ x\in \bb{F}_2^n \ |\  x\cdot c = 0 \text{ for all } c \sim \cal{C} \}$. A code is \emph{self-dual} if its dual code is itself.
\end{definition}
Here, the inner product is over $\bb{F}_2$. It is clear that $\dim \cal{C} + \dim \cal{C}^\perp = n$. The dimension of a self-dual code is $n/2$ with $n$ even.

We collect some results from classical coding theory in Lemma~\ref{lemma: QRC}.

\begin{lemma}\label{lemma: QRC}
The dimension of QRC with respect to $q$ is $(q + 1)/2$. Append a single bit to every codeword of QRC (these bits are not necessarily the same for all codewords), to make them all even. The extended QRC is self-dual and doubly even.
\end{lemma}

\subsubsection{The Quantum Value of Bias}

We append columns whose 1-st component is 0 to $\mathcal{X}_M$, to make a larger matrix $\mathcal{X}$. Obviously, columns in $\mathcal{X}_M$ are not orthogonal to $s = (1,0,0,0,0)$, but the appending columns are. We interpret $\cal{X}$ as an X-program, run it and calculate the bias in the direction of $s$, then we have the following theorem~\cite{IQP08}:
\begin{theorem}\label{thm_1}
Denote the code generated by $\cal{X}_M$ as $\cal{C}_{\rm QRC}$. Then the bias in the direction of $s$ is
\begin{align}\label{thm:quantum_value_of_bias}
\mathcal{P}_{s \bot } = \frac{1}{2^d} \sum_{c \in \cal{C}_{\rm QRC}} \cos^2[\theta (q - 2|c|)] \ ,
\end{align}
where $d = (q+1)/2$ is the dimension of QRC and $2^d$ is the number of elements in QRC.
% \begin{align}
% \beta = \Pr(x\cdot s = 0) = \mathbb{E}_{c\sim \mathcal{C}_s} [\cos^2(\theta (q - 2\cdot wt(c)))] \ .
% \end{align}
\end{theorem}
From the above formula, we know that the bias depends on the QRC $\cal{C}_{\rm QRC}$ instead of the matrix $\cal{X}_M$ or $\cal{X}$. This means that we can append $\cal{X}_M$ with arbitrarily many rows orthogonal to $s$, without changing the the bias. Also, we can do row manipulation to $\cal{X}$, and the bias remains unchanged as long as we change $s$ correspondingly s.t. the appending rows are still orthogonal to $s$. This can scramble the circuit and the secret vector in order to hide it. We write a code for generating the initial QRC matrix and scrambling, and put it at the end of this section as well as in Ref.~\cite{gitlab}.

From Lemma \ref{lemma: QRC}, we know that odd codewords in the original QRC have weight -1 modulo 4, and those of even parity have weight 0 modulo 4, since the extended QRC is doubly even. For a code, the number of even and odd codewords are equal. Thus $q - 2|c|$ in the summation is half the time 1 modulo 8, and half the time -1 modulo 8. So the bias (for $\theta = \pi/8$) is
\begin{align*}
\mathcal{P}_{s \bot } = \frac{1}{2} \cos^2{\frac{\pi}{8}} + \frac{1}{2} \cos^2{\left(-\frac{\pi}{8} \right) } = \cos^2{\frac{\pi}{8}} = 0.854 \ .
\end{align*}
This is the quantum value of bias.

\subsubsection{The Optimal Classical Value of Bias}

In \cite{IQP08}, a classical efficient algorithm was proposed to approximate the bias. The algorithm works as follows:
\begin{enumerate}
\item Randomly pick 2 $n$-bit vectors $d$ and $e$. (Recall that $n$ is the number of qubits as well as the length of columns in $\cal{X}$.)
\item Delete columns in $\cal{X}$ that are orthogonal to $d$, and denote the remaining rows as a matrix $\cal{X}_d$. Do the same to get $\cal{X}_e$.
\item Denote the sum of rows in $\cal{X}_d \cap \cal{X}_e$ as $y$. Then $y$ is the approximate result of $x$.
\end{enumerate}
By some simple calculation, we can show that $\Pr(y \ |\ y\cdot s = 0) = \Pr( c_1, c_2 \in \mathcal{C}_{\rm QRC} \ |\ c_1 \cdot c_2 = 0 )$~\cite{IQP08}. From Lemma \ref{lemma: QRC}, we know that the extended QRC is self-dual, which implies the even codewords in the extended QRC is orthogonal to all codewords. Since the extended QRC is obtained by appending a single bit to codewords in QRC to make them all even, the bit appended to even codewords in QRC is $0$. So the even codewords in QRC is orthogonal to all codewords in QRC. 
% the even codewords in $\cal{C}_{\rm QRC}$ is orthogonal to all codewords. 
If the inner product between two codewords is not zero, then they are all odd, which occurs with probability $1/4$. Thus
\begin{align*}
\Pr(y \ |\ y\cdot s = 0) = \Pr( c_1, c_2 \in \mathcal{C}_{\rm QRC} \ |\ c_1 \cdot c_2 = 0 ) = 1 - \frac14 = \frac34 \ .
\end{align*}
This value is conjected to be the best that a classical computer can approximate \emph{efficiently}.

\subsubsection{Python Code for QRC Construction}

% \begin{figure}[h]
\begin{minipage}{.48\textwidth}
\begin{lstlisting}[language=Python]
import numpy as np

# set parameters, which can be changed by readers
q = 7 # q must satisfy 8 divides (q+1)
n = int((q+3)/2) # number of qubits
r = 3 # number of redundant rows

def ColAdd(A, i, j):
    '''
    add the j-th column of A to the i-th column
    '''
    if len(A.shape) == 1:
        A_i = A[i]
        A_j = A[j]
        A_i = (A_i + A_j)%2
        A[i] = A_i
    else:
        A_i = A[:,i]
        A_j = A[:,j]
        A_i = (A_i + A_j)%2
        A[:,i] = A_i
    return A

def Redund(n):
    '''
    generate a redundant row
    '''
    a = np.random.choice(2, n)
    a[0] = 0
    return a

def QuadResidue(q):
    '''
    return quadratic residues modulo q
    '''
    qr = []
    for m in range(q):
        qr.append(m**2%q)
    qr.pop(0)
    return list(set(qr))

def Init(n, q, r):
    '''
    generate the matrix
    '''
    result = {}
\end{lstlisting}
\end{minipage}\hfill
\begin{minipage}{.48\textwidth}
\begin{lstlisting}[language = Python]
    P_s = np.zeros([q,n], dtype = int)
    P_s[:,0] = np.ones(q, dtype = int)
    qr = QuadResidue(q)
    for m in range(n-1):
        for m1 in qr:
            P_s[(m1-1+m)%q , m+1] = 1  

    P = P_s
    while r > 0:
        r -= 1
        row = Redund(n)
        if (row == np.zeros(n, dtype = int)).all():
            r += 1
            continue
        P = np.append(P, [row], axis = 0)
    result["matrix"] = np.unique(P, axis = 0)   # delete redundant rows
    
    s = np.zeros(n, dtype = int)
    s[0] = 1
    result["secret"] = s    
    return result

def Scramble(result, times):
    '''
    scramble P and s
    '''
    P = result["matrix"]
    s = result["secret"]
    for m in range(times):
        l = list(range(n))
        i = np.random.choice(l)
        l.pop(i)
        j = np.random.choice(l)
        P = ColAdd(P, i, j)
        s = ColAdd(s, j, i)
    result["matrix"] = P
    result["secret"] = s
    return result

result = Init(n, q, r) # generate the initial matrix
print(Scramble(result, 50)) # scramble 50 times, 
# and print the matrix 
# as well as the secret vector
\end{lstlisting}
\end{minipage}
% \caption{The python code to construct the $X$-program circuit run on \emph{ibmqx5}, with QISKit. }
% \label{fig:QRC_code}
% \end{figure}

\subsection{Probability Bias and $n$-point Correlation Functions}

In this section, we first show the relation between bias, Fourier components and $n$-point correlation functions. Then exploiting this relation, we show why the redundant part of $\mathcal{X}$ has no effect on the value of probability bias. Originally, in Ref.~\cite{IQP08}, this is proven through Theorem~\ref{thm_1}. Here, we provide a more straightforward and intuitive way to visualize this fact.

\begin{figure}
  \[
    \Qcircuit @C=1em @R=.7em {
    \lstick{\ket{0}} & \multigate{4}{\quad U \quad} & \gate{Z^{s_1}} & \multigate{4}{\quad U^\dagger \quad} & \rstick{\bra{0}} \qw \\
    \lstick{\ket{0}} & \ghost{\quad U \quad} & \gate{Z^{s_2}} & \ghost{\quad U^\dagger \quad} & \rstick{\bra{0}} \qw \\
    \lstick{\ket{0}} & \ghost{\quad U \quad} & \gate{Z^{s_3}} & \ghost{\quad U^\dagger \quad} & \rstick{\bra{0}} \qw \\
    \lstick{\ldots} & \ghost{\quad U \quad} & \push{\ \ldots\ } \qw & \ghost{\quad U^\dagger \quad} & \rstick{\ldots} \qw \\
    \lstick{\ket{0}} & \ghost{\quad U \quad} & \gate{Z^{s_n}} & \ghost{\quad U^\dagger \quad} & \rstick{\bra{0}} \qw 
    }
  \]
    \caption{Quantum circuit representation of $\langle Z^{s_1}\otimes Z^{s_2}\otimes \cdots \otimes Z^{s_n} \rangle$. Here, $U$ can be any quantum circuits. At the same time, this diagram can also be viewed as $\sum_x p(x) (-1)^{s\cdot x}$.}
    \label{fig:expectation}
\end{figure}
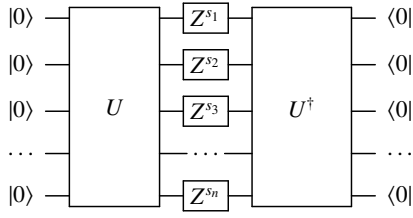

For a probability distribution $\{ p(x) \}$, its Fourier coefficient is 
\begin{align}
\hat{p}(s) &\equiv \frac{1}{2^n} \sum_x p(x) (-1)^{x\cdot s} \notag \\
&= \frac{1}{2^n} \left[ \sum_{x\cdot s = 0} p(x)  - \sum_{x\cdot s = 1} p(x) \right] \notag \\
&= \frac{1}{2^n} (2 \mathcal{P}_{s\perp} - 1) \ ,
\end{align}
from the normalization of $p(x)$. %Next, we show the relation between $n$-point correlation function and Fourier coefficients. 
Denote $\mathcal{Z}_s \equiv Z^{s_1} Z^{s_2} \cdots Z^{s_n}$, then the $n$-point correlation function for the final state $\rho$ is 
\[
\langle \mathcal{Z}_s \rangle \equiv \Tr(\mathcal{Z}_s \rho) = \mel{0^n}{U^{\dagger} (Z^{s_1} Z^{s_2} \cdots Z^{s_n}) U}{0^n} \ ,
\]
if $\rho$ is a pure state. Fig.~(\ref{fig:expectation}) shows the quantum circuit representation of $\langle \mathcal{Z}_s \rangle$. From this representation, we can see that it is actually the Fourier coefficient of $s$ (up to a normalization factor $1/2^n$):
\begin{align*}
    \ket{0^n} &\xrightarrow{U} \sum_x c_x \ket{x} \\
    &\xrightarrow{\mathcal{Z}_s} \sum_x c_x (-1)^{s\cdot x} \ket{x} \\
    &\xrightarrow{\bra{0^n} U^\dagger} \sum_x p(x) (-1)^{s\cdot x} \ ,
\end{align*}
where $c_x = \mel{x}{U}{0^n}$ is the transition amplitude. Thus $\hat{p}(s) = \langle \mathcal{Z}_s \rangle/2^n$, and 
\begin{align}\label{eq:bias_and_expectation}
\langle \mathcal{Z}_s \rangle = 2 \mathcal{P}_{s\perp} - 1 \ .
\end{align}
It should be noted that this relation is general and not restricted to IQP circuits.

An X-program circuit can be represented by a matrix $\mathcal{X}$ of binary values; its columns represent the gates, as in the Main Text. There are two parts in $\mathcal{X}$, the main part $\mathcal{X}_M$ and the redundant part $\mathcal{X}_R$. We also split the Hamiltonaian read from $\mathcal{X}$ into two part $H = H_M + H_R$, where $H_M$ is translated from $\mathcal{X}_M$ and $H_R$ is from $\mathcal{X}_R$. Then $U_{\rm IQP} = \exp(i\theta H) = e^{i\theta H_M} e^{i\theta H_R} $, since $H_M$ commutes with $H_R$. Columns in the redundant part of $\mathcal{X}$ are orthogonal to $s$, which implies that $H_R$ commutes with $\mathcal{Z}_s$ and so does $\exp(i\theta H_R)$.
For example, $(1, 1, 0, 0, 0)$ is orthogonal to $s = (1, 1, 1, 1, 0)$, and $\left[e^{i\theta X_1 X_2}, \mathcal{Z}_s \right] = 0$. As for $H_M$, it anticommutes with $\mathcal{Z}_s$, so $e^{i\theta H_M} \mathcal{Z}_s = \mathcal{Z}_s e^{-i\theta H_M}$. Thus 
\begin{align*}
\mel{0^n}{U^{\dagger}_{\rm IQP} \mathcal{Z}_s U_{\rm IQP}}{0^n} &= \mel{0^n}{e^{-i\theta (H_M + H_R)} \mathcal{Z}_s e^{i\theta (H_M + H_R)}}{0^n} \\
&= \mel{0^n}{e^{-i\theta H_M} \mathcal{Z}_s e^{i\theta H_M}}{0^n} \\
&= \mel{0^n}{e^{i2\theta H_M}}{0^n} \ ,
\end{align*}
which has no dependence on the redundant part. Together with Eq.~\eqref{eq:bias_and_expectation}, we can see that the value of probability bias $\cal{P}_{s\perp}$ does not depend on the redundant part.

\subsection{Noise effect on probability bias}

Consider a simple noise model: single-qubit bit-flip channels $\mathcal{E}(\rho) = (1 - \epsilon) \rho + \epsilon X\rho X$ applied to every qubits at the end of the circuits. This model is equivalent to the noise model described in the main text: dephasing channels on every qubit at each time step. Those dephasing channels in between commute with each other as well as the diagonal gates, so they accumulate and can be regarded as an effective single-qubit dephasing channel on every qubit. After commuting with the final layers of Hadamard gates, they become bit-flip channels. 
Below, we follow Ref.~\cite{Yung2017} to show that under this noise, $\langle \cal{Z}_s \rangle \to (1 - 2\epsilon)^{|s|} \langle \cal{Z}_s \rangle$, and hence from Eq.~\eqref{eq:bias_and_expectation}, ${\mathcal{P}_{s \bot }} \to \frac{1}{2}\left( {1 + {{\left( {1 - 2 \epsilon } \right)}^{\left| s \right|}}\left\langle \cal{Z}_s \right\rangle } \right)$.

First, the effect of the measurement operator is $\cal{M}(\rho) = \sum_x p(x) \Pi(x)$, where $\Pi(x) = \dyad{x}$. Note that $\cal{E}$ commutes with $\cal{M}$. To prove it, we shall first suppose $\rho$ is a single-qubit state. Then
\begin{align*}
\cal{M} (\cal{E}(\rho)) &= (1 - \epsilon) \cal{M}(\rho) + \epsilon \cal{M}(X\rho X) \\
&= (1 - \epsilon) \cal{M}(\rho) + \epsilon \left[p(0)\Pi(1) + p(1)\Pi(0) \right] \\
&= (1 - \epsilon) \cal{M}(\rho) + \epsilon X \cal{M}(\rho) X \\
&= \cal{E}(\cal{M}(\rho)) \ .
\end{align*}
Thus the effect of $\cal{E}$ can be viewed as flipping the measurement outcome with probability $\epsilon$. Note that in the expansion of $\cal{M}(\rho)$, $\Pi(x)$ is like $\ket{x}$, and they also form a linear space. In this linear space, we introduce a `Hadamard' operator $\cal{H}$: $ \cal{H}\Pi(i) \equiv \frac{1}{\sqrt{2}} \left[ \Pi(0) + (-1)^i \Pi(1) \right] $ for $i = 0, 1$. Then
\[
\cal{H}^{\otimes n} \Pi(x) = \frac{1}{\sqrt{2^n}} \sum_s (-1)^{x\cdot s} \Pi(s) \ .
\]
From the definition of $\cal{H}$, we can easily see that $\cal{H}^2 = I$. The original expansion of $\cal{M}(\rho)$ has probabilities as its coefficients. To explore the effect of noise on Fourier coefficiets, we might want to change basis from $\Pi(x)$ to $\cal{H}^{\otimes n} \Pi(s)$: 
\begin{align*}
\cal{H}^{\otimes n} \cal{M}(\rho) &= \sum_x p(x) \cal{H}^{\otimes n} \Pi(x) \\
&= \frac{1}{\sqrt{2^n}} \sum_x \sum_s p(x) (-1)^{x\cdot s} \Pi(s) \\
&= \sqrt{2^n} \sum_s \hat{p}(s) \Pi(s) \ , \\
\cal{M}(\rho) &= \cal{H}^{\otimes n} \cal{H}^{\otimes n} \cal{M}(\rho) \\
&= \sqrt{2^n} \sum_s \hat{p}(s) \cal{H}^{\otimes n} \Pi(s) \ .
\end{align*}
Thus in this new basis, the coefficient is exactly $\hat{p}(s)$ (up to a factor). Now, $\cal{E}\left[ \cal{H}\Pi(s_i) \right] = (1 - 2\epsilon)^{s_i} \cal{H}\Pi(s_i)$, which gives $ \cal{E}^{\otimes n}\left[ \cal{H}^{\otimes n}\Pi(s) \right] = (1 - 2\epsilon)^{|s|} \cal{H}^{\otimes n}\Pi(s) $, and hence $\hat{p}(s) \to (1 - 2\epsilon)^{|s|}\hat{p}(s)$. Since $\hat{p}(s) = \langle \cal{Z}_s \rangle / 2^n$, we have $\langle \cal{Z}_s \rangle \to (1 - 2\epsilon)^{|s|} \langle \cal{Z}_s \rangle$ under this noise model.

\subsection{Experimental Details}

\subsubsection{NMR processor}

\begin{figure}[t]
\includegraphics[width=0.9 \columnwidth]{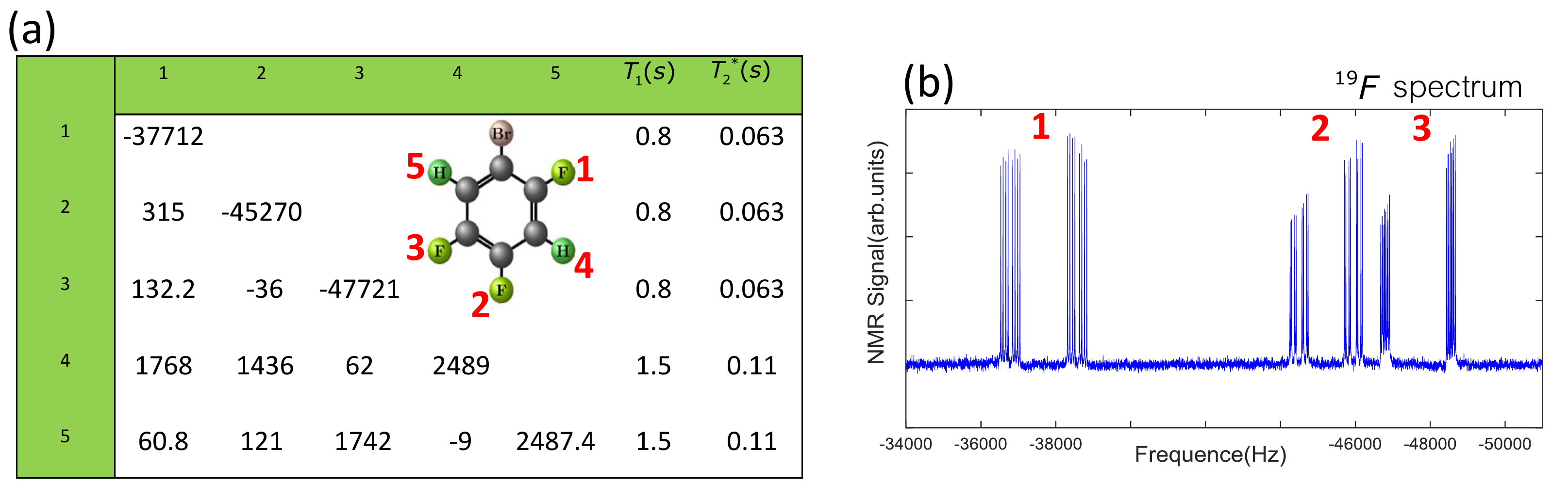}
\caption{(a) Molecular structure and parameter of 1-bromo-2,4,5-trifluorobenzene. The diagonal and off-diagonal elements represent the chemical shifts and effective coupling constants in units of Hz, respectively. (b) Equilibrium spectra of $^{19}F$ nuclear spins. The numbers above the peaks are the indices of qubits.
}\label{fig:sample}
\end{figure}

\emph{Device. }
The experiments run on NMR quantum processor were carried on a Bruker AV-400 spectrometer at 303K. The 5-qubit quantum processor consists of two ${}^1H$ nuclear spins and three ${}^{19}F$ nuclear spins in 1-bromo-2,4,5-trifluorobenzene dissolved in the liquid crystal N-(4-methoxybenzylidene)-4-butylaniline (MBBA)~\cite{SHANKAR201410-sample}. Figure~\ref{fig:sample} (a) shows the molecular structure and (b) shows equilibrium spectra of ${}^{19}F$ nuclear spins.

The effective Hamiltonian of the NMR sample in the rotating frame is denoted as:
\begin{align}
{H_{NMR}}{\text{ = }}\sum\limits_{j = 1}^5 \pi {\nu _j} Z_j  + \sum\limits_{1 \leqslant j \leqslant k \leqslant 5} {\frac{\pi }{2}} \left( {{J_{jk}} + 2{D_{jk}}} \right) Z_j Z_k \ ,
\label{eq:HNMR1}
\end{align}
where $Z_j$ are the Pauli-$Z$ matrices acting on the $j$-th qubit. The chemical shifts $\nu_j$ and the effective coupling constants ${{J_{jk}} + 2{D_{jk}}}$ are shown in Fig.~\ref{fig:sample} (a). Here we have employed the secular approximation since that the chemical shift difference in each pair of spins is much higher than the effective coupling strength.

\begin{figure}[t]
\includegraphics[width = 1\columnwidth]{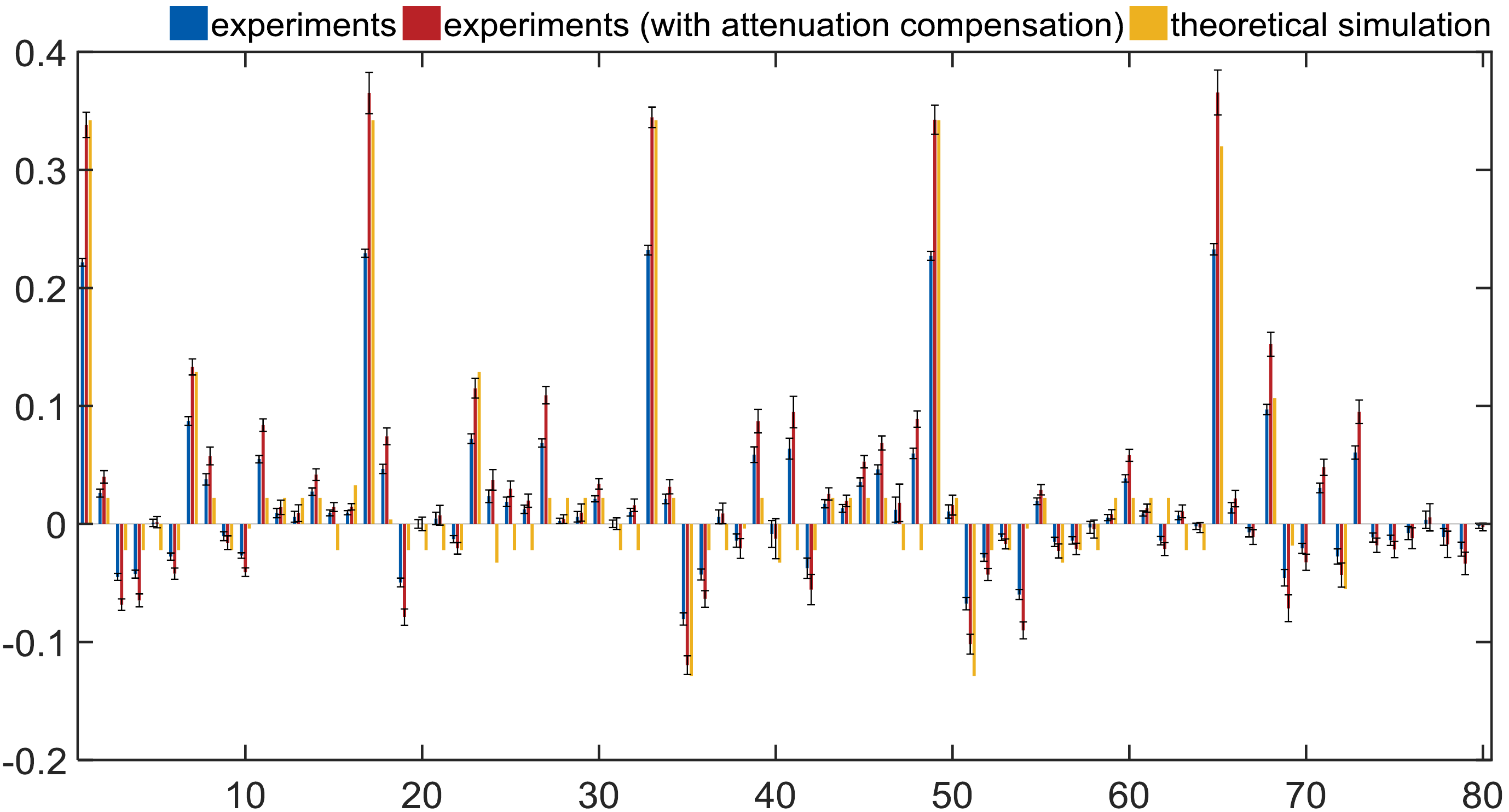}
\caption{Peak intensities from five readout pulses. Each pulse creates 16 peak intensities, which are linear combination of probabilities, so there are 80 peak intensities for each kinds of data. 
}
\label{fig:peak}
\end{figure}

\emph{Readout. } 
As we mentioned in the main text, the probability distributions for the IQP circuit is obtained by reconstructing the diagonal elements
of the density matrix of the final state. To obtain the information of the $j$-th qubit, We applied the readout pulse $\exp(-i\pi Y_j/4)$ and then read the NMR spectrum of it. The first qubit, i.e. $F_1$, is used as the readout channel in the experiment. The spectrum of $F_1$ is read directly and the information of other qubits are swapped to $F_1$ before obtaining the NMR spectra. For each qubit, we can obtain a readout spectrum which contains 16 peaks. Taking the spectrum of third qubit as example, the intensities of the peaks equal to $\Pr ({x_1}{x_2}0{x_4}{x_5}) - \Pr ({x_1}{x_2}1{x_4}{x_5})$, where
${x_1},{x_2},{x_4},{x_5} \in \{ 0,1\} $. From the intensities of all the 80 peaks,
We have 80 linear equations of the form: $\sum_{x} c_{lx} \Pr(x) = \alpha_l$ ($1\leqslant l\leqslant 80$), where $\alpha_l$ is the intensity of the $l$-th peak. Therefore, the probability distributions $\Pr(x)$ can be obtained by measuring the intensities of the peaks and then solving these linear equations.

\begin{figure}[t!]
    \centering
    \includegraphics[width = 0.9\columnwidth]{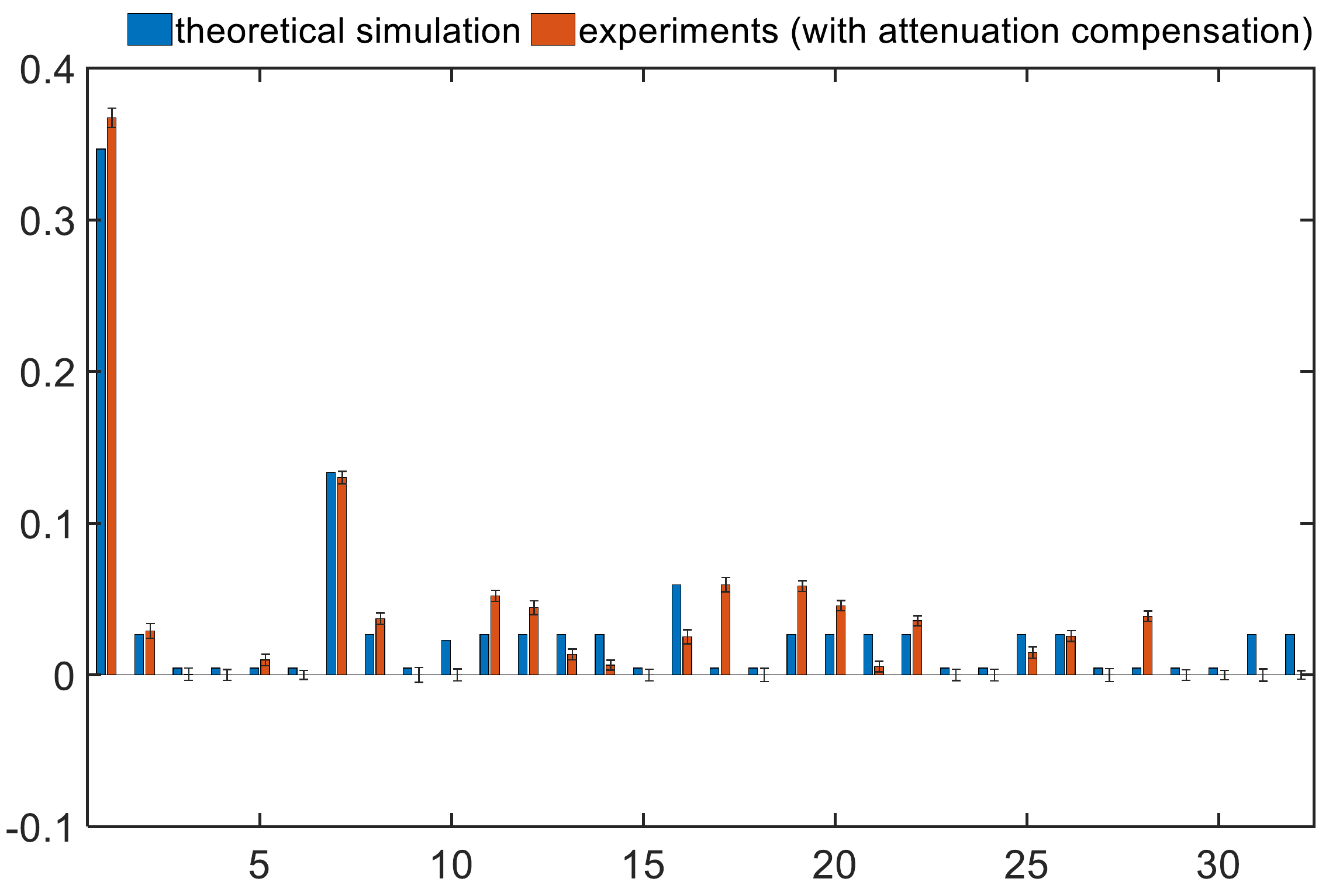}
    \caption{The blue histogram is from theoretical distribution and the red one is from distribution from the NMR device after noise compensation.}
    \label{fig:NMR_dist_compensation}
\end{figure}

The intensities of all the 80 peaks of the five spins are shown as the blue histogram in Fig.~\ref{fig:peak}. The orange one is the ideal result from the theoretical simulation. In the experiments, the intensities of the NMR signal suffer a significant attenuation due to the overlong lengths of the pulses. To compensate this effect, we numerically measure the attenuation speed of each spin by independent benchmarking experiments. The attenuation factors in the duration of the pulse sequence are estimated as $0.66 \pm 0.02$, $0.63 \pm 0.03$, $0.67 \pm 0.01$, $0.66 \pm 0.02$ and $0.64 \pm 0.03$ for the five spins. The attenuation of the decoherence effect then can be compensated by dividing the intensities of the peaks by the corresponding attenuation factors. In Fig.~\ref{fig:peak}, the red histogram shows the intensities of peaks with the attenuation compensation, which has a good agreement with the theoretical expectation. 

The probability distributions $\Pr(x)$ can be estimated by solving the overdetermined linear equations $\sum_{x} c_{lx} \Pr(x) = \alpha_l$ ($1\leqslant l\leqslant 80)$, on condition that $\sum_{x}\Pr(x) = 1$ and $\Pr (x) \geqslant 0$. Here $\alpha_l$ represents the compensated intensity of the $l$-th peak. With the least square method, we obtain a probability distribution after compensation of noise (Fig.~\ref{fig:NMR_dist_compensation}), whose probability bias with respect to the secret string $s$ is $0.866 \pm 0.016$. The distance between this distribution and the theoretical distribution is 0.23, where the distance between two distributions $p_i$ and $q_i$ is defined by $D = \frac{1}{2} \sum_{i=1}^{32} |p_i - q_i|$.

\emph{Estimation of the readout errors. }  
The readout errors of the intensities of the peaks mainly come from two parts: the statistical fluctuation of the NMR spectra and the error in the compensation procedure. In the experiments, the intensities of the peaks are obtained by fitting the NMR spectra to a sum of several Lorentz functions. The results of fitting indicate that these intensities have an average standard error of 0.004, shown as the error bars on the blue histogram in Fig.~\ref{fig:peak}. To compensate the attenuation of the signal due to the decoherence effect, these intensities are divided by an attenuation factor, e.g. $0.66 \pm 0.02$ for the peaks of the first spin. The error of the attenuation factors also contribute to the compensated results. The readout errors of the compensated intensities of the peaks are shown in Fig.~\ref{fig:peak}, with the average error being 0.007. The estimation of the readout errors in the probability distributions $\Pr(x)$ are shown as the error bars in Fig.~\ref{fig:NMR_dist_compensation}. The probability bias is estimated as 0.866, with a standard error of 0.016.

\begin{figure}[t]
\includegraphics[width=0.9 \columnwidth]{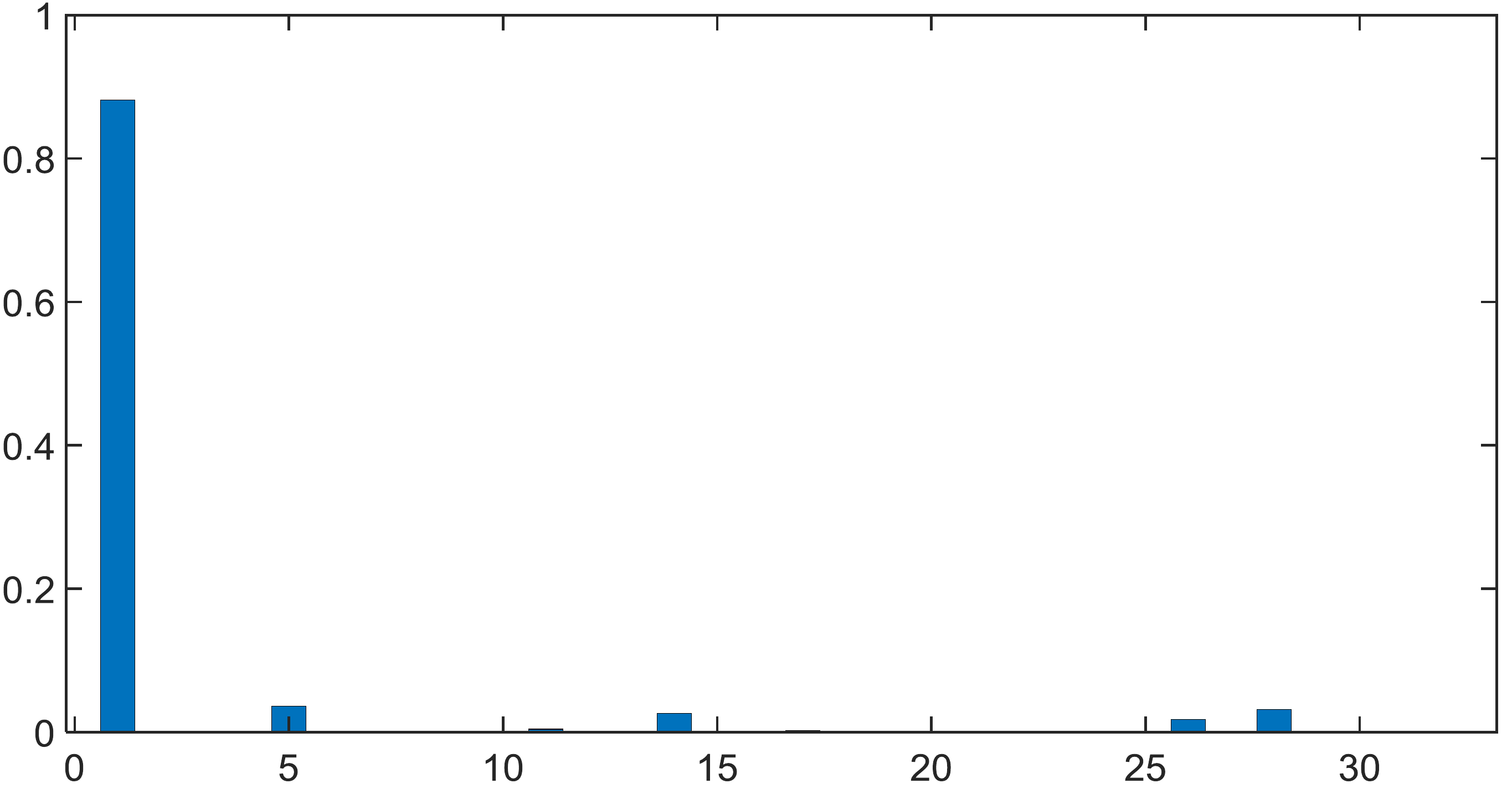}
\caption{The experimentally measured population of the initial state on the computational basis, listed in the order from $\left| {00000} \right\rangle $ to $\left| {11111} \right\rangle $.
}\label{fig:pps_pop}
\end{figure}

\begin{figure}[t]
\includegraphics[width=0.9 \columnwidth]{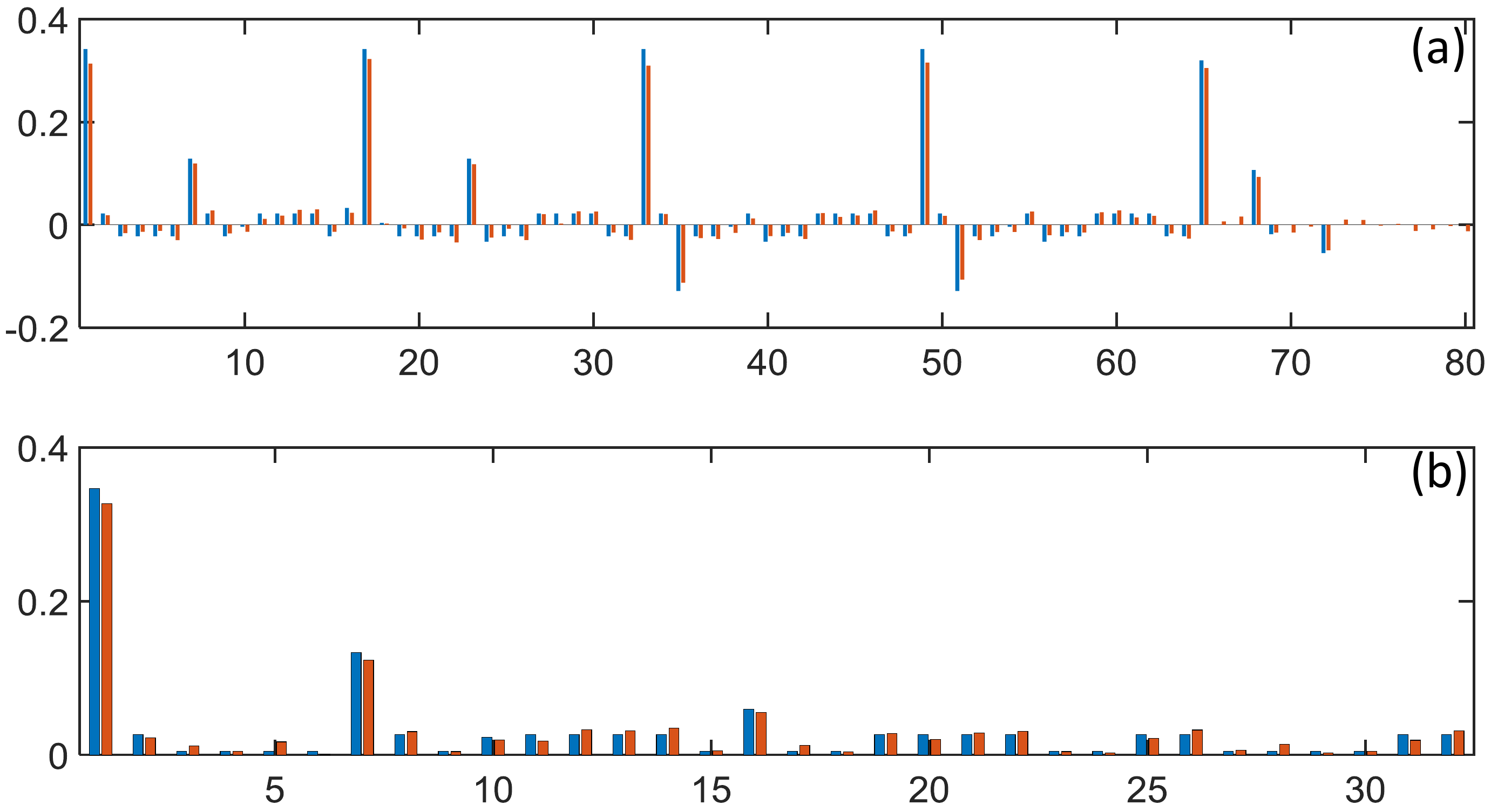}
\caption{The numerically simulated results of the final state with (red) and without (blue) the presence of the imperfection of the implementation.
(a) The intensities of the peaks of the readout spectra. (b) The probability distributions $\Pr(x)$ of the final state after the IQP circuit.
}\label{fig:error_all}
\end{figure}

\emph{The error from the imperfection of experimental implementation. }
In the experiments, the results are also affected by the imperfection of the implementation, including the imperfection of the initial state and the pulse sequences. We experimentally measured the population of the initial state on the computational basis (Fig.~\ref{fig:pps_pop}), while the ideal initial state is (pseudo) $\left| {00000} \right\rangle $ state. It shows that the population of some quantum state is not zero as expected and thus may affect the final results. The pulse sequence employed to realize the IQP circuit is also not perfect in the experiment. Numerical simulation shows that the fidelity of the shaped pulse used may reduce to 97.5\% if we take into consideration the imperfection of the secular approximation in the Hamiltonian of the quantum processor (Eq. \eqref{eq:HNMR1}) and the thermal fluctuation of the Hamiltonian parameters. To understand to what extent these imperfections will affect the experimental result, we numerically simulate these effects and show the final state with the presence of these imperfection in Fig.~\ref{fig:error_all}, as well as the ideal results without any error. The distance $D$ between ideal final state and the one with error is around 0.08. The probability bias is reduced to 0.832, while the ideal expectation is 0.854. Thus we estimated that these imperfections may lead to a drift of -0.02 on the probability bias.

\subsubsection{IBM Devices}

\begin{figure}[t]
    \centering
    \includegraphics[width = 0.9\columnwidth ]{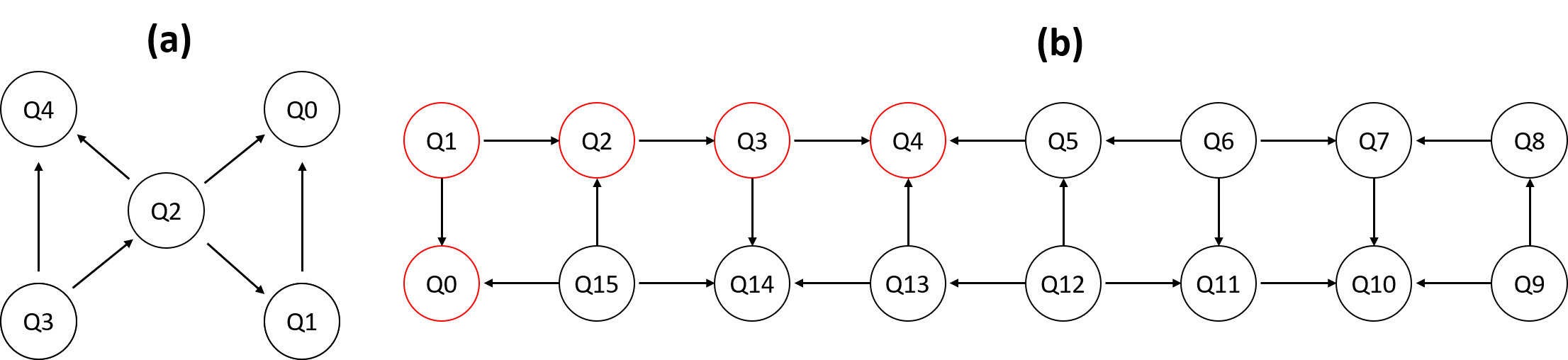}
    \caption{Connectivity of \textbf{(a)} \emph{ibmqx4} and \textbf{(b)} \emph{ibmqx5}. Here $a \to b$ means qubit $a$ controls qubit $b$. The red circles in \textbf{(b)} are the qubits we use in our experiment. }
    \label{fig:connectivity}
\end{figure}

We use \emph{ibmqx4} and \emph{ibmqx5} for our experiments. We collect some specification of IBM devices here, and readers can find more details in Ref.~\cite{ibmqx}. Fig.~\ref{fig:connectivity} shows the connectivity of the two IBM devices. There are 16 qubits in \emph{ibmqx5}, and we used the first 5 qubits, indicated as the red circles in Fig.~\ref{fig:connectivity} (b). The first table of Table~\ref{tab: ibm_specification} shows the frequency, relaxation time and coherence time of \emph{ibmqx4}, and the second shows those of \emph{ibmqx5}. The IBM devices are calibrated constantly, so the $T_1$ and $T_2$ listed in Table~\ref{tab: ibm_specification} are only mean values over an interval, according to the version number in Ref.~\cite{ibmqx}. The number after the $\pm$ sign shows the standard deviation of the mean.

\begin{figure}[t]
    \centering
    \includegraphics[width = 0.9\columnwidth ]{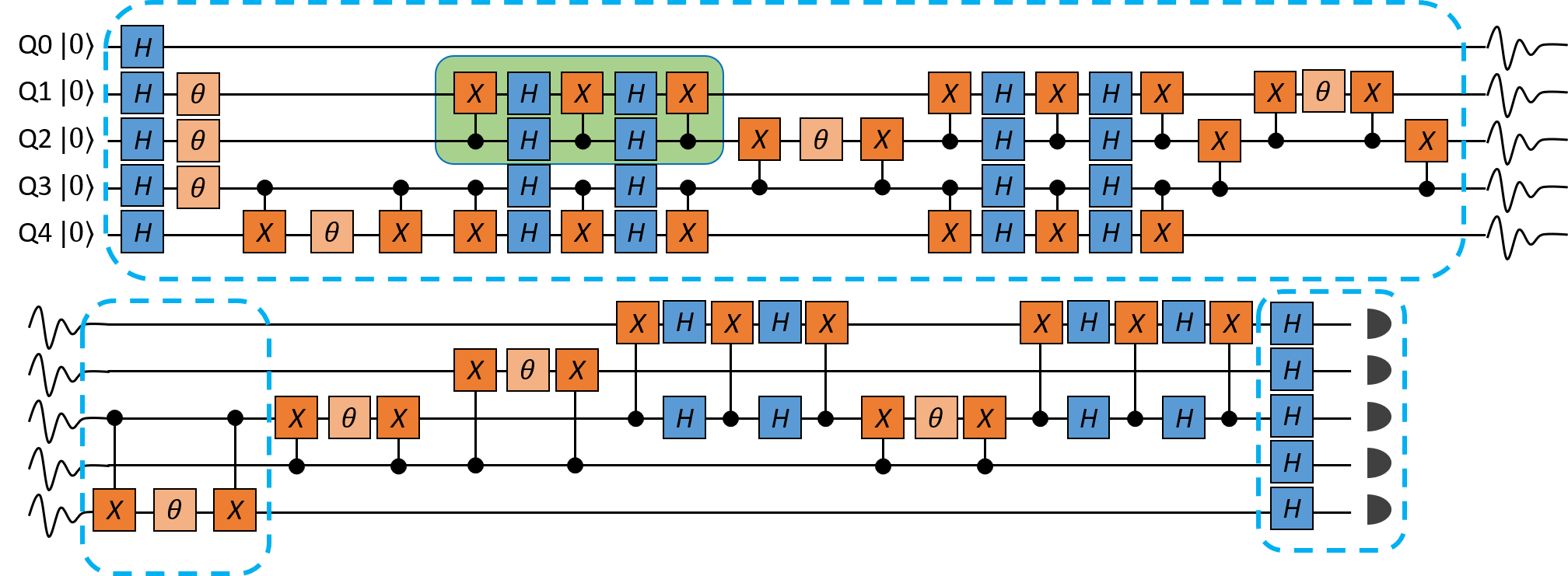}
    \caption{The $X$-program circuit run on \emph{ibmqx4}. The green block is a SWAP gate. Components enclosed by blue dashed lines form a circuit translated from the main part of the $X$-program matrix in the main text, i.e. matrix~\eqref{eq:main}. }
    \label{fig:ibmqx4_circuit}
\end{figure}

We access \emph{ibmqx4} through the website of IBM quantum experence and can edit the circuit on a graphical interface. Fig.~\ref{fig:ibmqx4_circuit} shows the circuit run on \emph{ibmqx4}. This circuit is equivalent to that of Fig.~2 in the main text. but due to the connectivity constraints on the device, we need to add many SWAP gates to implement the protocol on \emph{ibmqx4}. But for \emph{ibmqx5}, we need to access it through QISKit, an open-source software based on python. The code used to construct the circuit is shown in the end of the Supplemental Materials, and the full version can be found in Ref.~\cite{gitlab}.

\begin{table}%[h]
\begin{minipage}{0.9\textwidth}
\centering
\begin{tabular}[t]{|c|c|c|c|c|c|}
\hline
     & Q0               & Q1               & Q2               & Q3               & Q4               \\ \hline
\begin{tabular}[c]{@{}l@{}}Frequency \\ (GHz)\end{tabular} & 5.24 & 5.30 & 5.35 & 5.43 & 5.18 \\ \hline
$T_1(\mu s)$ & 48.81 $\pm$ 0.68 & 50.24 $\pm$ 0.66 & 42.52 $\pm$ 0.51 & 40.09 $\pm$ 0.94 & 55.52 $\pm$ 0.96 \\ \hline
$T_1(\mu s)$ & 28.09 $\pm$ 0.41 & 60.24 $\pm$ 1.12 & 34.92 $\pm$ 0.51 & 14.24 $\pm$ 0.21 & 27.09 $\pm$ 0.37 \\ \hline
\end{tabular}
% \captionof{table}{Frequency, relaxation time ($T_1$) and coherence time ($T_2$) of \emph{ibmqx4}.}
\end{minipage}
\vspace{5mm}

\begin{minipage}{0.9\linewidth}
\centering
\scalebox{0.9}{
\begin{tabular}{|l|l|l|l|l|l|l|l|l|l|l|l|l|l|l|l|l|}
\hline
   & Q0         & Q1          & Q2         & Q3          & Q4          & Q5         & Q6          & Q7         & Q8          & Q9          & Q10         & Q11         & Q12        & Q13         & Q14        & Q15         \\ \hline
\begin{tabular}[c]{@{}l@{}}Frequency \\ (GHz)\end{tabular} & 5.26 & 5.40 & 5.28 & 5.08 & 4.98 & 5.15 & 5.31 & 5.25 & 5.12 & 5.16 & 5.04 & 5.11 & 4.95 & 5.09 & 4.87 & 5.11 \\ \hline
$T_1(\mu s)$ & 37 $\pm$ 4 & 35 $\pm$ 4  & 48 $\pm$ 6 & 46 $\pm$ 5  & 49 $\pm$ 8  & 49 $\pm$ 4 & 44 $\pm$ 7  & 37 $\pm$ 4 & 49 $\pm$ 7  & 48 $\pm$ 5  & 28 $\pm$ 21 & 45 $\pm$ 7  & 50 $\pm$ 7 & 48 $\pm$ 6  & 36 $\pm$ 3 & 48 $\pm$ 7  \\ \hline
$T_2(\mu s)$ & 31 $\pm$ 5 & 58 $\pm$ 10 & 64 $\pm$ 7 & 70 $\pm$ 15 & 74 $\pm$ 24 & 50 $\pm$ 5 & 74 $\pm$ 12 & 49 $\pm$ 7 & 68 $\pm$ 19 & 88 $\pm$ 14 & 49 $\pm$ 36 & 86 $\pm$ 16 & 33 $\pm$ 3 & 82 $\pm$ 12 & 65 $\pm$ 6 & 89 $\pm$ 17 \\ \hline
\end{tabular}
}
% \captionof{table}{Frequency, relaxation time ($T_1$) and coherence time ($T_2$) of \emph{ibmqx5}.}
\end{minipage}
\caption{Specification of IBM devices. The first table collects specification of \emph{ibmqx4}, and the second collects that of \emph{ibmqx5}.}\label{tab: ibm_specification}
\end{table}

\begin{figure}[h!]
    \centering
    \includegraphics[width = 0.9\columnwidth ]{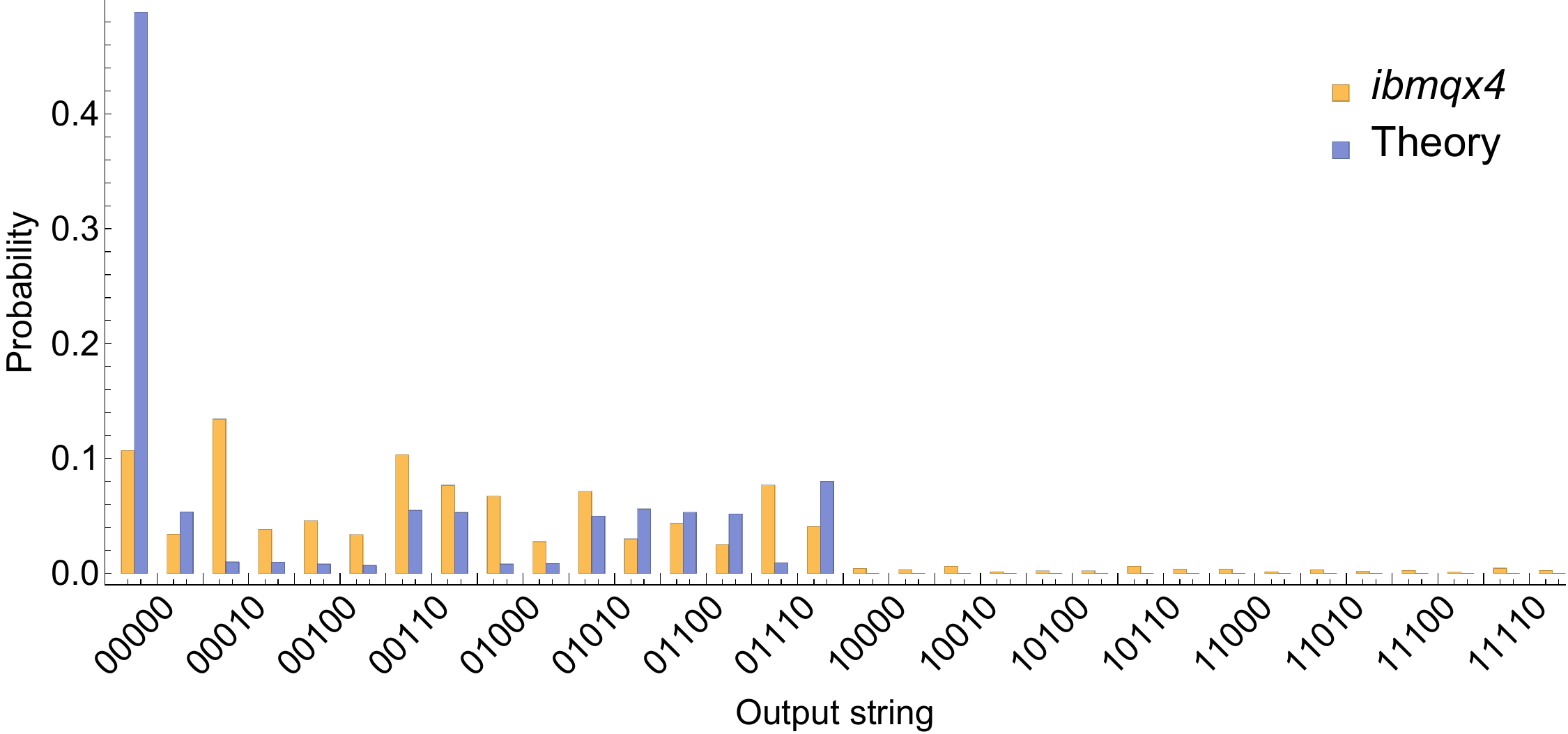}
    \caption{Probability distribution from the IQP circuit translated from matrix~\eqref{eq:main}.}
    \label{fig:ibmqx4_main}
\end{figure}

On \emph{ibmqx4}, we also implement the quantum circuit corresponding to the main part of the QRC matrix only, that is matrix~\eqref{eq:main}. Components in Fig.~\ref{fig:ibmqx4_circuit} enclosed by blue dashed lines represent the circuit. The output probability distribution is shown in Fig.~\ref{fig:ibmqx4_main}. As we show in the main text, the probability bias of the secret vector $s=(1,1,1,1,0)$ only depends on the main part of the $X$-program matrix. From the output distribution, we obtain that the bias is 0.512, which is still very close to bias derived from a completely mixed state. This indicates that even if we significantly reduce the depth of the circuit, the final state is still highly corrupted by noise. Thus, we conclude that the fidelity of IBM cloud needs to be significantly improved in order to pass the test.

\clearpage

\newpage

\textbf{Python code to specify the circuit run on \emph{ibmqx5}}
% Below is the python code to construct the $X$-program circuit run on \emph{ibmqx5}, with QISKit. 

% \begin{figure}[h!]
\begin{minipage}{.48\textwidth}
\begin{lstlisting}[language=Python]
# circuit

for i in range(5):
    qc.h(qr[i])  # Hadamard gate
    
for i in [1,2,3]:
    qc.tdg(qr[i])  # T-dagger
    
# (0,0,0,1,1)
qc.cx(qr[3], qr[4])  # CNOT with Q3 controling Q4
qc.tdg(qr[4])
qc.cx(qr[3], qr[4])   

# (0,1,0,0,1)
qc.cx(qr[1], qr[2])
qc.cx(qr[3], qr[4])
for i in [1,2,3,4]:
    qc.h(qr[i])
qc.cx(qr[1], qr[2])
qc.cx(qr[3], qr[4])
for i in [1,2,3,4]:
    qc.h(qr[i])
qc.cx(qr[1], qr[2])
qc.cx(qr[3], qr[4])

qc.cx(qr[2], qr[3])
qc.tdg(qr[3])
qc.cx(qr[2], qr[3])

qc.cx(qr[1], qr[2])
qc.cx(qr[3], qr[4])
for i in [1,2,3,4]:
    qc.h(qr[i])
qc.cx(qr[1], qr[2])
qc.cx(qr[3], qr[4])
for i in [1,2,3,4]:
    qc.h(qr[i])
qc.cx(qr[1], qr[2])
qc.cx(qr[3], qr[4])

# (0,1,1,1,0)
qc.cx(qr[1], qr[2])
qc.cx(qr[2], qr[3])
qc.tdg(qr[3])
qc.cx(qr[2], qr[3])
qc.cx(qr[1], qr[2])

# (0,0,1,0,1)
qc.cx(qr[3], qr[4])
for i in [3,4]:
    qc.h(qr[i])
qc.cx(qr[3], qr[4])
for i in [3,4]:
    qc.h(qr[i])
qc.cx(qr[3], qr[4])

qc.cx(qr[2], qr[3])
qc.tdg(qr[3])
qc.cx(qr[2], qr[3])

qc.cx(qr[3], qr[4])
for i in [3,4]:
    qc.h(qr[i])
\end{lstlisting}
\end{minipage}\hfill
\begin{minipage}{.48\textwidth}
\begin{lstlisting}[language = Python]
qc.cx(qr[3], qr[4])
for i in [3,4]:
    qc.h(qr[i])
qc.cx(qr[3], qr[4])

# (0,0,1,1,0)
qc.cx(qr[2], qr[3])
qc.tdg(qr[3])
qc.cx(qr[2], qr[3])

# (1,0,1,0,0)
qc.cx(qr[1], qr[0])
for i in [0,1]:
    qc.h(qr[i])
qc.cx(qr[1], qr[0])
for i in [0,1]:
    qc.h(qr[i])
qc.cx(qr[1], qr[0])

qc.cx(qr[1], qr[2])
qc.tdg(qr[2])
qc.cx(qr[1], qr[2])

qc.cx(qr[1], qr[0])
for i in [0,1]:
    qc.h(qr[i])
qc.cx(qr[1], qr[0])
for i in [0,1]:
    qc.h(qr[i])
qc.cx(qr[1], qr[0])

# (1,0,0,1,0)
qc.cx(qr[1], qr[0])
qc.cx(qr[2], qr[3])
for i in [0,1,2,3]:
    qc.h(qr[i])
qc.cx(qr[1], qr[0])
qc.cx(qr[2], qr[3])
for i in [0,1,2,3]:
    qc.h(qr[i])
qc.cx(qr[1], qr[0])
qc.cx(qr[2], qr[3])

qc.cx(qr[1], qr[2])
qc.tdg(qr[2])
qc.cx(qr[1], qr[2])

qc.cx(qr[1], qr[0])
qc.cx(qr[2], qr[3])
for i in [0,1,2,3]:
    qc.h(qr[i])
qc.cx(qr[1], qr[0])
qc.cx(qr[2], qr[3])
for i in [0,1,2,3]:
    qc.h(qr[i])
qc.cx(qr[1], qr[0])
qc.cx(qr[2], qr[3])

for i in range(5):
    qc.h(qr[i])

# measure
for j in range(5):
    qc.measure(qr[j], cr[j])
\end{lstlisting}
\end{minipage}
% \caption{The python code to construct the $X$-program circuit run on \emph{ibmqx5}, with QISKit. }
% \label{fig:ibmqx5_circuit}
% \end{figure}

% \clearpage

% \newpage

% %\nocite{*}
% % \bibliographystyle{unsrt} % full title in ref
% \bibliography{ref}

\end{document}